\documentclass[aps,final,a4paper,twocolumn,prb,10pt]{revtex4}%
\usepackage{amsfonts}
\usepackage{amsmath}
\usepackage{amssymb}
\usepackage[dvips]{graphicx}%
\providecommand{\U}[1]{\protect\rule{.1in}{.1in}}
\begin{document}
\title{Chern-number spin Hamiltonians for magnetic nano-clusters by
DFT methods}
\author{T. O. Strandberg and C. M. Canali}
\affiliation{Division of Physics,
Department of Natural Sciences,
Kalmar University, 391 82 Kalmar, Sweden}
\altaffiliation{Solid State Theory, Lund University, Lund, Sweden.}

\author{A. H. MacDonald}
\affiliation{Department of Physics,
University of Texas at Austin, Austin, Texas 78712, USA}

\pacs{PACS numbers(s): }

\begin{abstract}
Combining field-theoretical methods and ab-initio calculations, we
construct an effective Hamiltonian with a single giant-spin degree
of freedom, capable of the describing the low-energy spin dynamics
of ferromagnetic metal nanoclusters consisting of up to a few tens
of atoms. In our procedure, the magnetic moment direction of the
Kohn-Sham SDFT wave-function is constrained by means of a penalty
functional, allowing us to explore the entire parameter space of
directions, and to extract the magnetic anisotropy energy and
Berry curvature functionals. The average of the Berry curvature
over all magnetization directions is a Chern number - a
topological invariant that can only take on values equal to
multiples of one half, representing the dimension of the Hilbert
space of the effective spin system. The spin Hamiltonian is
obtained by quantizing the classical anisotropy energy functional,
after performing a change of variables to a constant Berry
curvature space. The purpose of this article is to examine the
impact of the topological effect from the Berry curvature on the
low-energy total-spin-system dynamics. To this end, we study small
transition metal clusters: Co$_{n}$ ($n=2,...,5$), Rh$_{2}$,
Ni$_{2}$, Pd$_{2}$, Mn$_{x}$N$_{y}$, Co$_{3}$Fe$_{2}$.

\end{abstract}
\volumeyear{2007}
\volumenumber{Volume number}
\issuenumber{Issue number}
\eid{identifier}
\date{070717}
\received[Received text]{date}

\revised[Revised text]{date}

\accepted[Accepted text]{date}

\published[Published text]{date}

\maketitle

\section{Introduction}
The present interest in magnetic nanoparticles, nano-clusters
and molecular magnets\cite{skomski_book,molmagnet}
is partly motivated by the quest
for nano-scale information
storage devices\cite{murray_science2000},
and by possible applications in nano-spintronics
and quantum computation devices. Most of the advances achieved in the last few
years in this field have been fueled by the remarkable experimental ability to
synthesize, manipulate and characterize
these systems at the atomic scale\cite{skomski_book}.
Theoretical work, based on refined
models\cite{pastor2003, ac_cmc_ahm2002, skomski_nanomag03} and
advanced computational tools\cite{ebert05, kashyap}, is also
providing crucial understanding and new predictions.
Magnetic metal nano-clusters,
like other magnetic nanostructures, typically display enhanced
magnetic properties\cite{billas1994} and many novel classical and
quantum phenomena,
which challenge our understanding
of magnetism\cite{deHeer_prl05,chelikowsky_prl06}.

A quantity that is the topic of intense interest in nanomagnetism,
is the magnetic anisotropy. This is the dependence of the total
energy of the system on the orientation direction of the magnetic
moment. The presence of two or more stable orientations in the
anisotropy energy functional, separated by an energy barrier, is
the crucial property exploited in information storage devices. The
origin of Magnetocrystalline Anisotropy Energy (MAE) in magnetic
materials is subtle, being a relatively small effect arising
primarily from the spin-orbit interaction. It is now possible to
investigate experimentally the MAE in individual magnetic
nanoparticles\cite{jamet2001}
 and nano-clusters\cite{gambarella03Sci}
with unprecedented accuracy. Typically the MAE in these magnetic
nanostructures is found to be larger than in bulk. Gambardella
{\it et al.}\cite{gambarella02Nat1d} found that monoatomic Co
chains on a stepped Pt(111) surface are ferromagnetic and have a
MAE that is one-order of magnitude larger than bulk Co.
Understanding the reasons of this enhancement and how to increase
it further is one of the goals of ongoing research. Recent
theoretical work predicts that the MAE in unsupported 4d
transition metal (TM) chains\cite{mokrousov_prl06} and
dimers\cite{tos_cmc_ahm_natmat2007} have giant MAE up to 50 meV.

Another topic of great importance, connected to the MAE and
equally intensively investigated, is the spin
dynamics of the collective magnetization orientation degree of freedom
in single-domain nanoparticles. For a nanoparticle containing several thousand
atoms, the magnetic orientation degree of freedom is a classical variable that
oscillates around one energy minimum of the MAE or switches between two of
them by thermal activation. Thermal switching in {\it individual}
superparamagnetic
nanoparticle has been recently studied by spin-polarized
scanning tunneling microscope\cite{bode_prl04}. When the nanocluster is
sufficiently small, the quantum behavior of the collective spin becomes
important. The crossover from classical to quantum behavior for
the magnetization collective variable is a complex problem of
great scientific interest. The physical systems on which research has
focused so far in studying
this question are molecular
magnets in a crystal\cite{qtm94,chudnovsky_tejada98,molmagnet}.
These systems contain about 100 atoms per particle, are insulators and
behave like a single microscopic quantum spin, e.g.
$S=10$ for Mn$_{12}$ acetate.
Quantum effects are clearly discernible in their low-temperature
relaxation properties.  A large amount of literature has been devoted to
studying these systems\cite{molmagnet}.
Similar studies of the quantum spin dynamics in
engineered magnetic metal nanoclusters have just begun.
In a recent key experiment,
Hirjibehedin {\it et al.}\cite{Hirjibehedin_sci06:_MnCuN} investigated
collective spin
excitations in {\it individual} scanning tunnel microscope (STM)-engineered
linear chains of 1 to 10 manganese atoms,
by means of inelastic electron tunneling spectroscopy. Spin excitations that
change both the total collective spin and the spin orientation were observed.

On the theoretical front, the microscopic description of quantum spin
excitations in magnetic metal clusters is more challenging than in
insulating molecular magnets, since in the metal clusters
these excitations are intertwined with electronic
particle-hole excitations. The need to include the crucial
spin-orbit interactions adds to the complexity of the task.
However, in a paper
published a few years ago\cite{ccmPRL03},
two of us put forward the idea that, for ultra-small
metal clusters, where the single-particle energy level spacing is typically
larger
than then total MAE barrier, the "fast" (high-energy) electronic
degrees of freedom can be integrated out. It is then
still possible to describe the
``slow'' (low-energy) dynamics of the collective magnetization
orientation in terms of
a single quantum mechanical
{\em giant spin} $J$ only, like in molecular magnets.
When spin-orbit interactions are
neglected, the ground state of a magnetic nano-cluster will tend to have a large
total spin quantum number $S$, and may have additional orbital degeneracies $g_{orb}$ as well.
In the atomic limit for example, the ground state is characterized by total spin and
total orbital angular momentum $L$ quantum numbers so that $g_{orb}=2L+1$.  In clusters, orbital
symmetries are reduced and $g_{orb}$ will tend to be small - in the simplest cases equal to $1$.
In this case, we identify the {\em giant spin} $J$  with $S$.
When spin-orbit interactions are included, this large ground state degeneracy will be lifted.
Following Ref.~\onlinecite{ccmPRL03}, the approach we take in this paper
assumes that the magnetic nano-cluster has a well-defined
group of low-lying energy levels which can be identified with the magnetic moment
orientation degree-of-freedom.
For a nano-cluster with total giant spin $J$, $2J+1$ many-body states will lie in this group.
For atoms, for example, Hunds third rule states that $J=L \pm S$.
The spectrum of these levels is described by a {\em giant spin} low-energy effective
Hamiltonian which is the quantum generalization of the magnetic anisotropy energy.

When integrated out, the fast electronic degrees of freedom produce a Berry
phase term in the effective action that can profoundly modify the behavior of
the magnetization orientation degree of freedom. Indeed in this approach,
orbital contributions, arising from spin-orbit interactions, are automatically
included via the Berry phase term. The giant-spin of the formalism is a
topological half-integer invariant known as Berry phase Chern number, which
characterizes the topologically non-trivial dependence of the nano-cluster's
many-electron wave functions on magnetization orientation.

The purpose of this paper is to carry out the procedure suggested in
Ref.~\onlinecite{ccmPRL03} by calculating Berry phase Chern numbers and
extracting effective spin Hamiltonians for a variety of realistic magnetic
nanoclusters containing different transition metals. In order to carry out this
procedure, we calculate the MAE for all these systems. We treat the many-body
electronic problem using state-of-the-art numerical methods based on
Spin-Density Functional Theory (SDFT). For transition metals having the largest
MAE such as Cobalt, our procedure is in principle valid for clusters containing
up to hundred atoms. However, with current state-of-the-art plane wave SDFT
codes and the advantages of parallel processing, the computational load sets
our estimated upper limit at a few tens of atoms. Here we investigate
unsupported ultra-small clusters with no more than 5 TM atoms. We neglect
non-collinear spin configurations and limit ourselves to coherent magnetization
solution, although we allow the possibility of both ferromagnetic and
antiferromagnetic order parameters. We focus, in particular, on the challenging
situation, quite ubiquitous in metal clusters, where the HOMO-LUMO gap
vanishes. This is the case where the Berry phase contribution is crucial. By
comparing our treatment with the ordinary procedure of extracting spin
Hamiltonians from the MAE functional, where the Berry phase term is neglected,
we show that the latter can lead to incorrect results for the thickness of the
magnetic anisotropy energy and for the oscillation frequencies of the
magnetization orientation degree of freedom.

The paper is organized as follows.
In section II we give a short description of the theoretical background and
the technical details needed to understand the output of the model. We then
present our main results in section III and proceed with a more in-depth
analysis of the different systems in section IV. Finally, we summarize our
conclusions in section V.

\section{Theory}

\subsection{The Berry curvature and the Chern number}

The fundamental origin of the Berry phase \cite{originalberry_prs84} is some
parametrization of a given term in the Hamiltonian, representing a coupling to
the external world\footnote{By {\it coupling to the external world}, it is meant
either a coupling to an external classical system or
a coupling to another quantum
system. The case we are interested is in fact the latter, where the quantum
electronic system is coupled to the collective magnetization degree
of freedom.}
It is this parametrization that leads to an
observable which cannot be cast into the form of an Hermitian operator, but is
instead given in terms of a phase change in the systems wavefunction.
In the systems we are interested in, this parameter is the orientation
of the coherent magnetization direction of the magnetic cluster, represented
by the unit vector $\hat{n}$.
The dependence of the many-body wavefunction on the unit vector
$\hat{n}$ gives rise to the Berry phase.
In its most fundamental
form, the phase change in the ground state wavefunction as we vary the
magnetization between points $1$ and $2$ is given by%
\begin{equation}
\Delta\varphi_{12}=-\text{Im}\log\langle\Psi(\hat{n}_{1})|\Psi(\hat{n}%
_{2})\rangle\equiv-\text{Im}\log\langle1|2\rangle,
\end{equation}
which is 
an arbitrary phase that can be gauged away. By closing the loop in traversing
the vertices of a triangle we obtain a non-vanishing phase change:
$-$Im$~\log\langle1|2\rangle\langle2|3\rangle \langle3|1\rangle$. In the
continuum limit of a discrete, closed path $\gamma$
we obtain the usual definition of the Berry phase%
\begin{equation}
\mathcal{P}=%
i{\displaystyle\oint\nolimits_{\gamma}}
d\hat{n}\cdot\langle\Psi|\mathbf{\nabla}_{\hat{n}}\Psi\rangle.
\end{equation}
If we then convert the line integral to a surface integral over
the enclosed area, we can identify the integrand as a very
convenient, gauge invariant quantity and define the Berry
curvature
\cite{resta2000, auerbach94:_inter_elect_quant_magnet},%
\begin{equation}
\mathcal{\vec C}[\hat{n}]=i\mathbf{\nabla}_{\hat{n}}\times\langle\Psi
|\mathbf{\nabla}_{\hat{n}}\Psi\rangle. \label{curv}%
\end{equation}
The dimensionality of the total spin system is encoded in the dependence of
the many-body wavefunction on the magnetization direction as a topological
invariant called the Chern number\cite{qian_niu_book},
obtained by taking the average of the Berry
curvature over the unit sphere\cite{ccmPRL03},
\begin{equation}
\mathcal{S}=\frac{1}{4\pi}\int_{S^{2}}
\mathcal{\vec C}[\hat{n}]\cdot\hat{n}\; dA. \label{cnum}%
\end{equation}
Below we will refer to $\mathcal{C}[\hat{n}]
\equiv \mathcal{\vec C}[\hat{n}]\cdot \hat n $
as the {\it Berry curvature}.
Without SOI, the $\mathcal{C}[\hat{n}]$ is constant and is trivially
equal to the
total spin quantum number
$S$ \cite{cmc_ac_ahm2002pap4}. There is no anisotropy as the magnetization
direction is varied. Anisotropy comes from the coupling of electronic spin and
orbital degrees of freedom via the inclusion of SOI. The dimension of the
Hilbert space where the systems total effective spin resides is now the good
quantum number and total angular momentum $J$, replacing $S$ as the Chern
number of the system. For example, consider the case of a single Co atom:
disregarding SOI, the system has a total spin of $S=\tfrac{3}{2}$ coming from
the $3d$ electrons. Including SOI, we instead find a Chern number equal to
$J=\frac{9}{2}$, in accordance with the ground state predicted by Hund's
rules. Of course there is no anisotropy for a single atom as it is
rotationally invariant in the absence of a molecular or crystal field.

When the size and geometrical complexity of the cluster increases,
determination of the total $J$ quantum number quickly becomes a
highly non-trivial task. As is commonly the case, one may assume
that SOI does not cause a level-crossing at the HOMO level. The
dimensionality of the system can then be inferred from the total
magnetic moment in the absence of SOI, after which the spin-orbit
shifts can be estimated perturbationally. Adding the caveat of a
dense level structure at the HOMO level, as in small transition
metal clusters, there is the possibility that a spin-orbit induced
level-crossing may alter the dimensionality, rendering the
aforementioned reasoning specious. In the presence of SOI, one may
again attempt to infer the total spin dimension from the magnetic
moment, but this is a risky process as the numerical difficulties
associated with its determination for TM clusters can easily cause
a round-off error. By contrast, the Chern number can only take on
values of multiples of one-half, resolving any ambiguity in
obtaining the dimension of the total spin system.

It is useful to consider the perturbational expression for the Berry curvature.
In the so-called
parallel transport gauge\cite{resta2000} the phase of the wavefunction is kept
constant as the magnetization direction is infinitesimally varied.
Each infinitesimally translated state then becomes orthogonal to the previous state.
This means that the wavefunction can become multiply valued and obtain a sign
change as $\hat{n}$ executes a closed loop on the unit sphere. The choice of parallel transport gauge is implicitly made in the perturbational expression,
\begin{equation}
|\frac{\partial\Psi^{0}}{\partial n_{i}}\rangle=\sum_{m\neq0}|\frac
{\partial\Psi^{m}}{\partial n_{i}}\rangle\frac{\langle\Psi^{m}|\frac{\partial
H\left(  \hat{n}\right)  }{\partial n_{i}}|\Psi^{0}\rangle}{E_{0}\left(
\hat{n}\right)  -E_{m}\left(  \hat{n}\right)  }, \label{pert1}%
\end{equation}
where we denote the ground state by $\Psi^{0}$ and the sum runs over
excited states $\Psi^{m}.$ (\ref{pert1}) can be inserted into the expression
for the curvature (\ref{curv}) to yield%
\begin{equation}
\hat{\mathcal{C}}[\hat{n}]=i\sum_{i=1}^{3}\sum_{m\neq0}\frac{\langle\Psi^{0}%
|\frac{\partial H\left(  \hat{n}\right)  }{\partial n_{j}}|\Psi^{m}%
\rangle\langle\Psi^{m}|\frac{\partial H\left(  \hat{n}\right)  }{\partial
n_{k}}|\Psi^{0}\rangle}{\left[  E_{0}\left(  \hat{n}\right)  -E_{m}\left(
\hat{n}\right)  \right]  ^{2}}\hat n_i\;\varepsilon_{ijk}. \label{pertcurv}%
\end{equation}
Although this equation does no service in calculating the curvature (one would
have to calculate the excited states), it clearly exhibits the strong
dependence of the curvature on the HOMO-LUMO gap.

\subsection{Quantum Hamiltonian Extraction}
\label{quant_ham}

We follow the procedure outlined in \cite{ccmPRL03} and start from
an approximate imaginary-time quantum action that is a functional of the
magnetization direction $\hat{n}$ of the single effective spin degree of
freedom,%
\begin{equation}
S[\hat{n}]\equiv\int d\tau\left[  \langle\Psi\lbrack\hat{n}]|\mathbf{\nabla
}_{\hat{n}}\Psi\lbrack\hat{n}]\rangle\cdot \frac{\partial\hat{n}}{\partial\tau
}+E[\hat{n}]\right]\; . \label{qaction}%
\end{equation}
The orientational degree of freedom $\hat n$ is more precisely
defined as the direction in spin-space of the SDFT
exchange-correlation effective field\cite{uhl1994}, or its spatial
average if low-energy states have non-collinear magnetization.
In Eq.~(\ref{qaction}) $\vert \Psi [\hat n] \rangle$ is the
Kohn-Sham single-Slater determinant ground state defined by this
orientation, and $E[\hat n]$ is the SDFT energy. Here spin-orbit
terms are included in the Hamiltonian. For simple model
Hamiltonians where the exchange interaction can be decoupled by
means of auxiliary field functional integrals, Eq.~(\ref{qaction})
can be derived microscopically, starting with a path integral over
electronic coherent states. Eq.~(\ref{qaction}) was originally
proposed by Niu and Kleinman\cite{niu1998,niu1999} to study the
spin-wave dynamics of bulk TM ferromagnets in the context of the
adiabatic approximation, and successfully exploited in the actual
derivation of magnon dispersion by means of density functional
theory\cite{gebauer2000,bylander2000}.

The Euler-Lagrange equations for extremizing the action
Eq.~(\ref{qaction}) yield the semiclassical dynamics of the
orientation direction $\hat n$,
\begin{equation}
\mathcal -i C_{\alpha\beta}[\hat n]\,{\dot n}_{\beta} +
\frac{\partial E}{\partial n_{\alpha}} =0\;,
\end{equation}
where the Berry curvature matrix elements $\mathcal C_{\alpha\beta}$
are defined as
\begin{equation}
\mathcal C_{\alpha\beta}[\hat n] =
\frac{\partial}{\partial n_{\alpha}}
\langle \Psi\lbrack\hat{n}]|\frac{i \partial}{\partial \hat n_{\beta}}|\Psi\lbrack\hat{n}] \rangle -
\frac{\partial}{\partial n_{\beta}}
\langle \Psi\lbrack\hat{n}]|\frac{i \partial}{\partial \hat n_{\alpha}}|\Psi\lbrack\hat{n}] \rangle\;.
\end{equation}
These equations reduce to the classical Landau-Lifshitz (LL)
equations when the curvature is constant and equal to the total
spin moment $S$. In the presence of SOI however, the curvature is
in general non-constant and can deviate considerably from $S$,
particularly along directions in correspondence of which  the
HOMO-LUMO energy gap is small. Thus the LL equations can fail in
this case. In this paper, we aimed instead at deriving a quantum
mechanical description of the effective total spin, whose
magnitude we identify with the Chern number $\mathcal S$. Equipped
with the magnetic anisotropy energy and the Berry curvature
functionals on the unit sphere, we can proceed with the extraction
of an effective quantum Hamiltonian for this spin. Assuming that
the curvature is positive definite, the low energy effective
action (\ref{qaction}) can be mapped onto that of an effective
quantum spin Hamiltonian by first performing a variable
transformation to constant curvature space. This Hamiltonian is a
convenient tool that can be used for calculations of tunneling
amplitudes, non-linear response to external fields
etc. The variable transformation is given by\cite{ccmPRL03}%
\begin{align}
\phi &  \rightarrow\phi^{\prime}=\frac{2\pi\int_{0}^{\phi}\mathcal{C}\left(
u,\phi^{\prime\prime}\right)  d\phi^{\prime\prime}}{\int_{0}^{2\pi}%
\mathcal{C}\left(  u,\phi^{\prime\prime}\right)  d\phi^{\prime\prime}%
},\label{trphi}\\
u  &  \rightarrow u^{\prime}=-1+\frac{1}{2\pi\mathcal{S}}\int_{-1}%
^{u}du^{\prime\prime}\int_{0}^{2\pi}d\phi^{\prime\prime}\mathcal{C}\left(
u^{\prime\prime},\phi^{\prime\prime}\right)  . \label{trtheta}%
\end{align}
where $\mathcal{C}$ is the curvature, $\mathcal{S}$ the Chern number and
$u=\cos\theta.$ This change of variables rescales the local curvature metric
such that%
\begin{equation}
\mathcal{C}\left(  u,\phi\right)  dud\phi=\mathcal{S}du^{\prime}d\phi^{\prime
}. \label{metric}%
\end{equation}
From this expression we see that the transformed infinitesimal area element
will be rescaled in order to compensate for variations in the curvature.
Consequently, if $\mathcal{C}>\mathcal{S}$ the variable transform will cause
an expansion of the grid points and if $\mathcal{C}<\mathcal{S}$ a
contraction. At each grid point there is an energy value associated with the
magnetization direction that will be relocated by the transform. In
particular, we see from expression (\ref{pertcurv}) that if the HOMO and the
LUMO level become quasi-degenerate for one or more magnetization directions,
the curvature will become quasi-singular, causing a substantial deformation of
the grid.

The variable transformation allows us to rewrite the real-time action
as
\begin{equation}
\mathcal{S}_{\text{spin}}[\hat{n}^{\prime}]\equiv\int_{0}^{t}dt^{\prime
}\left[  i\;\vec{A}\cdot\frac{d\hat{n}^{\prime}}{dt^{\prime}}-E\left[  \hat
{n}\left(  \hat{n}^{\prime}\left(  t^{\prime}\right)  \right)  \right]
\right]  ,
\end{equation}
where $\vec{A}=s\hat{\phi}^{\prime}\left(  1-\cos\theta^{\prime}\right)
/\sin\theta^{\prime}$. This is the quantum action for a total spin of quantum
number $s$, with a \emph{classical} Hamiltonian $H\left[  \hat{n}^{\prime
}\left(  t^{\prime}\right)  \right]  \equiv E\left[  \hat{n}\left(  \hat
{n}^{\prime}\left(  t^{\prime}\right)  \right)  \right]  $
\cite{auerbach94:_inter_elect_quant_magnet}, that is equal to the expectation
value of a \emph{quantum} spin Hamiltonian $\mathcal{H}$,%
\begin{equation}
H\left[  \hat{n}^{\prime}\left(  t^{\prime}\right)  \right]  \equiv\langle
s,\hat{n}^{\prime}\left(  t^{\prime}\right)  |\mathcal{H}|s,\hat{n}^{\prime
}\left(  t^{\prime}\right)  \rangle, \label{clasqm}%
\end{equation}
with respect to the spin coherent states $|s,\hat{n}^{\prime}\left(
t^{\prime}\right)  \rangle$ describing a spin $s$ parametrized by the unit
vector $\hat{n}^{\prime}\left(  t^{\prime}\right)  $.

The quantization of the transformed magnetic anisotropy energy is then
performed by expanding in spherical harmonics of highest possible order
$2s$, where time-reversal symmetry precludes odd-powered harmonics from
contributing.
\begin{equation}
E\left[  \hat{n}\left(  \hat{n}^{\prime}\left(  t^{\prime}\right)  \right)
\right]  =\sum_{l=0}^{2s}\sum_{m=-l}^{l}\gamma_{l}^{m}Y_{l}^{m}\left(  \hat
{n}^{\prime}\right)  \label{classexp}%
\end{equation}
The spin coherent state $|s,\hat{n}^{\prime}\left(  t^{\prime
}\right)  \rangle$ can be constructed by applying rotation operators in the
single spin-space acting on the coherent spin state with maximum
polarization $m=s$. This allows us to express
the RH side of (\ref{clasqm}) in terms of spherical harmonics.%
\begin{align}
&  \sum_{mm^{\prime}}\langle s,s|\mathcal{D}^{\dag}\left(  R\right)
|s,m\rangle\langle s,m|\mathcal{H}|s,m^{\prime}\rangle\langle s,m^{\prime
}|\mathcal{D}\left(  R\right)  |s,s\rangle\nonumber\\
&  =\sum_{mm^{\prime}}\left(  R_{sm}^{s}\right)  ^{\dag}\mathcal{H}%
_{mm^{\prime}}^{s}R_{m^{\prime}s}^{s}\\
&  =\sum_{mm^{\prime}l\tilde{m}}\left(  -1\right)  ^{m-s}\sqrt{4\pi\left(
2l+1\right)  }\nonumber\\
&  \times\left(
\begin{array}
[c]{ccc}%
s & s & l\\
-s & s & 0
\end{array}
\right)  \left(
\begin{array}
[c]{ccc}%
s & s & l\\
-m & m^{\prime} & \tilde{m}%
\end{array}
\right)  Y_{l}^{\tilde{m}}\left(  \hat{n}^{\prime}\right)  \mathcal{H}%
_{mm^{\prime}}^{s} \label{formula}%
\end{align}
where we have made use of the Wigner-Eckhart theorem and expressed the
Clebsch-Gordan coefficients in terms of Wigner-3J symbols. Equating
(\ref{formula}) to the RH side of (\ref{classexp}), convoluting
with $Y_{\lambda}^{\mu}$ and inverting, we obtain%
\begin{align}
H_{mm^{\prime }}^{s} & =\left( -1\right) ^{m^{\prime }-s}\sum_{\lambda \mu
}\gamma _{\lambda }^{\mu }\sqrt{\frac{2\lambda +1}{4\pi }} \nonumber\\
&  \times\left(
\begin{array}{ccc}
s & s & \lambda  \\
m & -m^{\prime } & \mu
\end{array}%
\right) \left/ \left(
\begin{array}{ccc}
s & s & \lambda  \\
s & -s & 0%
\end{array}%
\right) \right. \label{wigner}%
\end{align}
Equation (\ref{wigner}) is of great practical use as gives an analytical
relation between the spherical harmonic expansion coefficients and the matrix
elements of the quantum Hamiltonian. Once the Hamiltonian matrix has been
obtained, it can be decomposed in terms of the spin operators $S_{x},$
$S_{y},$ $S_{z}$ and $S^{2}$.

\subsection{Computational Details}

Our numerical tool is SDFT using the
Projector Augmented Waves (PAW) formalism \cite{originalpaw_prb94} as
implemented in the VASP \cite{vaspcode} code. As such, the method depends on
representing the atoms with PAW pseudopotentials, in which the core electrons
are described on an auxiliary radial grid. This method can be said to be of
all electron type, although only the valence electrons participate in the full
variational procedure, whereas the core region is adjusted by applying
boundary and conservation conditions. The pseudopotentials are generated in
the generalized gradient approximation which gives a better description of the
exchange-correlation effects.

We use accurate parameter settings for our calculations. By this we mean a
valence (auxiliary) energy cutoff of 348(621)$\allowbreak$, 298(488),
351(709), 326(541), 381(751), 351(740), 520(815) eV in the case of Co, Rh, Ni,
Pd, Fe, Mn and N, respectively. In that order, the PAW pseudopotentials
\cite{vasppaw_psps} for these elements are generated in the following
electronic valence configurations: $3d^{8}4s^{1}$, $4d^{8}5s^{1}$,
$3d^{9}4s^{1}$, $4d^{9}5s^{1}$, $3p^{6}3d^{7}4s^{1}$, $3p^{6}3d^{6}4s^{1}$ and
$2s^{2}2p^{3}$. To avoid wrap-around errors we chose a dense FFT-grid taking
into account all $\vec{G}$-vectors that are twice as large as the vectors in
the basis set for the valence region and the 16/3 times as large for the
auxiliary region. Interaction between the periodically repeated cells is
suppressed by selecting a relatively large supercell-size of $11\times
11\times11$ A (in the case of Mn$_{x}$N$_{y}$ we use elongated cells
suitable for these geometries). Only the $\Gamma$-point is
used in the calculations and we have monitored dispersion by comparing the
eigenenergy spectrum at the $\Gamma$-point to that of the X-point, at the edge
of the Brillouin zone - in general there is no or very little dispersion. By
making use of the Vosko-Wilk-Nusair \cite{vosko-wilk-nusair_param}
interpolation formula for the correlation part of the exchange-correlation
functional, magnetic moments and energies are enhanced. All calculations are
performed in non-collinear mode (as described in ref.~\onlinecite{noncol-vasp}) with and
without SOI.

The calculation of electronic properties for small TM clusters is
a daunting task. Well converged solutions are often obscured by a
dense level structure at the HOMO level, causing sporadic jumps
between different spin-configurations and in a faulty ordering of
the eigenlevels. As noted by Pederson et.~Al
\cite{pederson_prb91:Pfrac_occ} (as well as
\cite{castro_prb97:_smearquote_and_coclusters, smearquote2_ni}),
the use of a finite (in our case Gaussian) smearing makes it
possible to achieve convergence in the self-consistent state. This
fake temperature causes a decreasing fractional occupation of the
levels around the HOMO level, enabling the correct ordering of
eigenstates. At the end of the calculation, the limit of zero
smearing is taken, eliminating the artificial entropy contribution
to the total energy.

SOI is included and the pseudopotentials are generated in the scalar
relativistic approximation
\cite{Koelling_jpc77:_scal_rel_soi, macdonald_jpc80:_scal_rel_soi}.
In this approach the relativistic Dirac
equation is separated into two parts. The first part has a $j$-weighted/averaged
scalar-relativistic Hamiltonian where the dependence on $\kappa=\pm(j+1/2)$ has
been removed. The second part has a spin-orbit Hamiltonian,%

\begin{equation}
\mathcal{H}_{SO}=\frac{{\hbar}}{(2Mc)^{2}}\frac{1}{r}\frac{dV}{dr}(\vec{\ell
}\cdot\vec{s}\hspace{1mm}),
\end{equation}
where $\vec{s}$ and $\vec{\ell}$ are the spin and angular momentum
operators and the relativistically enhanced electron mass is given
by $M=m+(\epsilon-V)/2c^{2}$.
Here $V$ is the effective potential, $c$ the speed of light and $\epsilon$
the eigenenergy. The scalar-relativistic Hamiltonian contains the mass
velocity and Darwin corrections. It is solved by diagonalization, whereafter
the spin-orbit Hamiltonian is included, such that the full Hamiltonian is
solved using the scalar relativistic wave functions as basis set. This
approach gives spin-orbit shifts that are in excellent agreement with
experimental values, although for the heavier elements the $p$-electron
$j$-weighting/averaging procedure introduces an approximation error
\cite{soi-description}.

Our solution for a given magnetization direction is stabilized via the
introduction of a penalty functional in the SDFT Hamiltonian, the additional
total energy contribution of which can be made vanishingly small. This
procedure allows us to calculate the total ground state energy for any given
direction on the unit sphere. In addition, we obtain the Kohn-Sham
wavefunction corresponding to the particular magnetization direction, which
can be used to calculate overlaps between wavefunctions for different
magnetization directions, and subsequently extract the Berry curvature. The
phase obtained in moving around a closed loop $\gamma$ may be written in terms
of the occupied Kohn-Sham single particle orbitals $\varphi_{i}$ as
\cite{resta2000}%
\begin{equation}
\mathcal{P}=i\sum_{i=1}^{N}%
{\displaystyle\oint\limits_{\gamma}}
\langle\varphi_{i}(\hat{n})|\mathbf{\nabla}_{\hat{n}}\varphi_{i}(\hat
{n})\rangle dl. \label{kslap}%
\end{equation}
In practice, we cover the unit sphere with small triangles and discretize
(\ref{kslap}) to a closed path with three vertices, expressing it terms of the
determinant of the overlap matrices $O_{ij}$ between the Kohn-Sham
wavefunctions obtained at each calculation vertex,
\begin{equation}
\mathcal{P}_{i}=-\text{Im}\log\left(  \det O_{12}^{i}\det O_{23}^{i}\det
O_{31}^{i}\right)  . \label{discphase}%
\end{equation}
where the index $i$ labels the triangles. Dividing (\ref{discphase}) by the
subtended area, we obtain the local curvature values $\mathcal{C}_{i}$.
Averaging these over the unit sphere yields the Chern number, which
can only take on multiples of one-half.

\begin{figure}[ptb]
\resizebox{7cm}{!}{\includegraphics{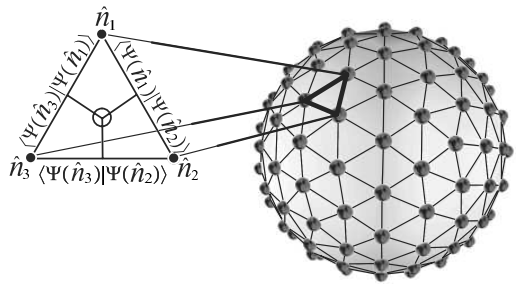}}\caption{The vertex skeleton
employed. At each vertex the ground state energy and the Kohn-Sham wavefunction
are obtained and for every triangle we may form the three overlaps needed to
calculate the curvature.}%
\label{sphere}%
\end{figure}

The computational load of this scheme is considerable and it is therefore
necessary to choose an optimal covering of calculation vertices on the unit
sphere. A convenient geometry is the 4-frequency icosahedron. An
$n$-frequency icosahedron is an icosahedron in which each of the 20 constituent
triangular surfaces has been subdivided into $n^{2}$ new triangles (see for
example ref.~\onlinecite{buckminsterfuller}).
Fig.~\ref{sphere} shows the 4-frequency icosahedron
with its 162 vertices, at which the ground state energies the associated Kohn-Sham wavefunctions are calculated, and 320 triangles, for which the Berry curvature
values are extracted.

\section{Results}

\subsection{Structural Optimization}

Structural optimization of the Co$_{n}$ ($n=2,...,5$), Ni$_{2}$ clusters is
carried out starting from initial atomic positions determined by Castro et.
Al. \cite{castro_cpl97:_fenico}, whereby the interatomic forces are minimized
and the structures refined using an accurate quasi-Newton method. Initial
distances for Pd$_{2}$, Rh$_{2}$ were taken from
ref.~\onlinecite{hafner_jpc05:_pd_rh}.
In accordance with the substrate-enforced geometry
(as in ref.~\onlinecite{Hirjibehedin_sci06:_MnCuN}), the Mn atoms are frozen
equidistantly at 3.6 A, with N in between. The Mn$_{x}$N$_{y}^{\ast}$
clusters are marked with an asterisk to indicate that the positions
have not been relaxed. We will also examine the effect of allowing
the atoms to relax their position on the axis.
For Mn$_{2}$N$_{3}$ we start relaxation from
an initial state where the N atoms are in plane but off the Mn axis.

\begin{figure}[ptb]
\resizebox{8cm}{!}{\includegraphics{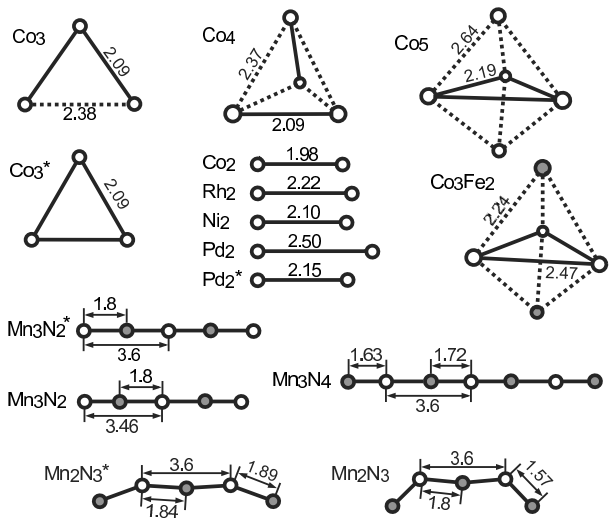}}\caption{The considered
clusters with distances in Angstrom. The asterisk indicates non-ground state
configurations. In the Mn$_{3}$N$_{4}^{*}$ cluster the atoms are equidistantly
spaced at 1.8 A, like in Mn$_{3}$N$_{2}^{*}$.
}%
\label{positions}%
\end{figure}

In addition we consider the equilateral Co trimer
(marked Co$_{3}^{\ast}$) and the effect of reducing the dimer distance in
Pd$_{2}$ (marked Pd$_{2}^{\ast}$). The considered structures are shown in
fig.~\ref{positions} where the distances are given in units of Angstrom. Table
\ref{bdesym} shows the resulting symmetries and binding energies.
\begin{table}
\begin{tabular}
[c]{|c||c|c|c|c|}\hline \emph{System} & \emph{Symmetry} & $E_{B}$
[eV] & $S$ & $J$ \\\hline Rh$_{2}$ & $C_{2h}$ & 3.68 & 2 &
4\\\hline Ni$_{2}$ & $C_{2h}$ & 2.87 & 1 & 1\\\hline Pd$_{2}$ &
$C_{2h}$ & 1.36 & 1 & 1\\\hline Pd$_{2}^{\ast}$ & $C_{2h}$ & 0.69
& 1 & 2\\\hline Co$_{2}$ & $C_{2h}$ & 3.57 & 2 & 4\\\hline
Co$_{3}$ & $C_{2h}$ & 6.46 & $7/2$ & $7/2$\\\hline Co$_{3}^{\ast}$
& $C_{3h}$ & 6.31 & $5/2$ & $3/2$\\\hline Co$_{4}$ &$D_{2d}$ &
10.01 & 4 & 4\\\hline Co$_{5}$ & $D_{3h}$ & 14.55 & $13/2$ &
$13/2$\\\hline Co$_{3}$Fe$_{2}$ & $D_{3h}$ & 14.22 & $13/2$ &
$11/2$\\\hline
Mn$_{2}$N$_{3}$ & $C_{2h}$ & 14.66 (14.69) & 5/2 (1/2) & 5/2 (1/2)
\\\hline
Mn$_{2}$N$_{3}^{*}$ & $C_{2h}$ & 12.39 (12.40) & 5/2 (1/2) & 5/2 (1/2) \\\hline
Mn$_{3}$N$_{2}$ & $C_{2h}$ & 12.83 (13.36) & 13/2 (5/2) & 13/2 (5/2) \\\hline
Mn$_{3}$N$_{2}^{*}$ & $C_{2h}$ & 12.67 (13.02) & 15/2 (5/2) & 15/2
(5/2)\\\hline Mn$_{3}$N$_{4}$ & $C_{2h}$ & 20.23 (20.06) & 9/2 (3/2)& 9/2 (3/2)
\\\hline Mn$_{3}$N$_{4}^{*}$ & $C_{2h}$ & 19.57 (19.36) & 9/2 (3/2) & 9/2
(3/2)\\\hline
\end{tabular}
\caption{The symmetries, binding energies and Chern numbers of the
considered clusters. The Chern numbers without ($S$) and with SOI
($J$) may differ in the presence of an SO-induced
level-crossing. Values in parenthesis refer to the AFM configuration.}%
\label{bdesym}%
\end{table}

\subsection{Chern Numbers}
The Chern numbers for the considered clusters, as obtained by
taking the average Berry curvature over the unit sphere, are
displayed in table \ref{bdesym}.
\begin{figure}[ptb]
\resizebox{8cm}{!}{\includegraphics{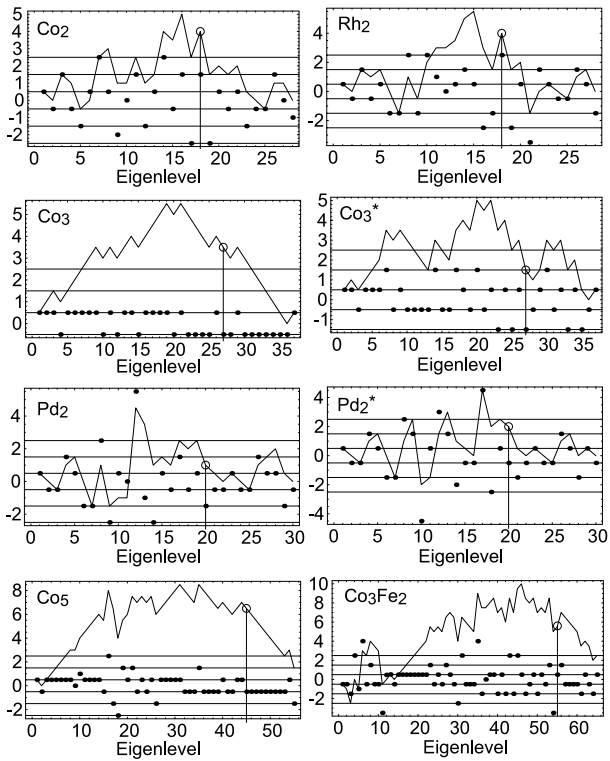}}\caption{The Chern
spectrum for a few selected clusters. The points show the level Chern numbers
and the solid line the accumulated Chern numbers. The HOMO level and the total
system Chern number have been marked with a vertical line and a circle.
Deviations from $\pm\frac12 $ indicate the presence of a level-crossing as a
function of increasing spin-orbit coupling strength. Multiples of half-integers
have been marked with horizontal lines in order to
lend support to the eye.}%
\label{cspec}%
\end{figure}

Fig.~\ref{cspec} shows the 'Chern spectrum' for a few chosen clusters. These
diagrams indicate the \emph{accumulated} Chern number (the solid line), obtained by
including the single-particle Kohn-Sham orbitals up to the given level in the
wavefunction overlaps, and the \emph{level} Chern number (the points), obtained by
taking the difference between accumulated Chern number of the given level and
that of its antecedent. We shall refer to the associated quantities as the
level and accumulated Chern number and curvature respectively.

In Fig.~\ref{cspec}, level Chern numbers are formed by an orbital part and a
spin part obtained by taking the average curvature over the unit
sphere. In the absence of SOI, the level Chern numbers can only take on values
of $\pm%
\frac12
$. As the spin-orbit coupling strength is increased from zero,
only levels that cross can deviate from this value by acquiring an
orbital part. In accordance with the Chern number sum rule for
level-crossings found in ref.~\onlinecite{ccmPRL03} - the sum of
the level Chern numbers before and after the crossing is
preserved. Notice that the level Chern numbers of the dimers are
particularly affected by level-crossings, attributable to their
inherent high degree of symmetry. Comparing the Jahn-Teller
distorted Co$_{3}$ with the equilateral trimer Co$_{3}^{\ast},$ we
see that the symmetrization of the
cluster leads to multiple deviations from $\pm%
\frac12
.$ A similar situation occurs in Co$_{5}$ when the two axial Co
atoms are substituted for Fe. The level-crossing effect may in
isolated cases even cause neighboring level Chern numbers to take
on integer values.

\begin{figure}[ptb]
\resizebox{8cm}{!}{\includegraphics{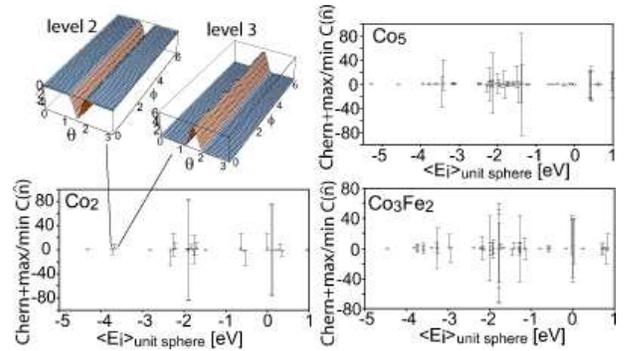}}\caption{ (Color online)
The Chern number plus the maximum and minimum level curvature as a function of
the average eigenenergy taken over the unit sphere. Note the large positive
maximum at the HOMO-level (set to zero) of Co$_{2}$ and the induction of
high-curvature variations obtained by substituting the axial Co atoms in
Co$_{5}$ with Fe.}%
\label{cdev}%
\end{figure}


In fig.~\ref{cdev}, we plot individual level Chern numbers
together with their corresponding maximum and minimum curvature
values on the unit-sphere as error bars, for a few representative
clusters. Large fluctuations in the level curvature signal the
presence of a near degeneracy between two adjacent levels. In
general we find that a level with large fluctuations in the Berry
curvature is followed by another level with curvature fluctuations
of the opposite sign that nearly compensate the previous ones.
(see for example the case of Co$_{2}$ where the curvature
landscape of two consecutive levels is shown.) This pattern
implies that although the Berry curvature of individual levels
might be singular in some directions, their sum is in general a
smooth function of $\hat n$ due to the cancellation of the
opposite contributions of adjacent levels. An exception to this
rule occurs when the quasi-degeneracy involves the HOMO and LUMO
levels. In this case, the fluctuations of the HOMO are not
compensated by the unoccupied LUMO; the \emph{total Berry
curvature} of the system is completely dominated by the HOMO
curvature and can therefore deviate strongly from the total Chern
number and be singular for particular magnetization directions. If
the total curvature becomes infinite, it may be difficult to
implement the procedure outlined in Sec.~\ref{quant_ham} . In
particular if the HOMO level has a predominantly minority-spin
character, the total curvature can even become negative along
these degeneracy directions, and the transformations given in
Eqs.~\ref{trphi}, \ref{trtheta} is invalid. Physically this means
that when HOMO-LUMO gap goes to zero the low-energy quantum
dynamics of the system cannot be described by an effective
giant-spin Hamiltonian only. In this case it necessary to include
explicitly all the electronic degrees of freedom involved in the
degeneracy at the Fermi level. We will address this issue in more
detail in Sec.~\ref{alt_app}. Depending on the given situation,
one might still to a good approximation evaluate the topological
effect when there are only isolated negative dips in the
curvature.
\begin{table}[ptb]%
\begin{tabular}
[c]{|l|l|l|}\hline
\emph{System} & $E_{MA}/atom$ [meV] & \emph{Type}%
\\\hline\hline
Rh$_{2}$ & \multicolumn{1}{|c|}{26} &
\multicolumn{1}{|c|}{EA}\\\hline Ni$_{2}$ &
\multicolumn{1}{|c|}{3.8} & \multicolumn{1}{|c|}{EP}\\\hline
Pd$_{2}$ & \multicolumn{1}{|c|}{1.2} &
\multicolumn{1}{|c|}{EP}\\\hline Pd$_{2}^{\ast}$ &
\multicolumn{1}{|c|}{17} & \multicolumn{1}{|c|}{EA}\\\hline
Co$_{2}$ & \multicolumn{1}{|c|}{14} &
\multicolumn{1}{|c|}{EA}\\\hline
Co$_{3}$ & \multicolumn{1}{|c|}{1.3/0.7} & \multicolumn{1}{|c|}{EA}%
\\\hline
Co$_{3}^{\ast}$ & \multicolumn{1}{|c|}{0.8} &
\multicolumn{1}{|c|}{EA}\\\hline Co$_{4}$ &
\multicolumn{1}{|c|}{0.9} & \multicolumn{1}{|c|}{EA}\\\hline
Co$_{5}$ & \multicolumn{1}{|c|}{0.1} &
\multicolumn{1}{|c|}{EA}\\\hline
Co$_{3}$Fe$_{2}$ & \multicolumn{1}{|c|}{0.5} & \multicolumn{1}{|c|}{EA}%
\\\hline
Mn$_{2}$N$_{3}$ & \multicolumn{1}{|c|}{0.9/0.4 (0.8/0.4)} &
\multicolumn{1}{|c|}{EA(EA)}\\\hline Mn$_{2}$N$_{3}^{*}$ &
\multicolumn{1}{|c|}{0.8/0.2 (0.7/0.1)} &
\multicolumn{1}{|c|}{EA(EA)}\\\hline Mn$_{3}$N$_{2}$  &
\multicolumn{1}{|c|}{1.4 (0.8)} &
\multicolumn{1}{|c|}{EP(EP)}\\\hline Mn$_{3}$N$_{2}^{\ast}$ &
\multicolumn{1}{|c|}{ 1.2 (1.0)} &
\multicolumn{1}{|c|}{EP(EA)}\\\hline Mn$_{3}$N$_{4}$  &
\multicolumn{1}{|c|}{ 0.8 (1.2)} &
\multicolumn{1}{|c|}{EP(EP)}\\\hline Mn$_{3}$N$_{4}^{\ast}$ &
\multicolumn{1}{|c|}{0.8 (0.3)} &
\multicolumn{1}{|c|}{EP(EA)}\\\hline
\end{tabular}
\caption{The magnetic anisotropy energies per atom. EA signifies
an easy axis in parallel to the symmetry axis of the cluster,
meaning the axis perpendicular to the atomic plane for
Co$_{3}^{\ast}$, the pyramid axis for Co$_{5}$ and
Co$_{3}$Fe$_{2}$, the axis running through the mid-point of the
two opposite solid lines in fig.~\ref{positions} for Co$_{4}$. EP
signifies a quasi-easy plane perpendicular to the symmetry axis
running through the dimers (or the line connecting the Mn atoms).
Co$_{3}$ has a high and a low barrier separating the global minima
that occur when the magnetization is parallel to the elongated
side. The low barrier runs in a cross-section perpendicular to the
trimer plane, whereas the high barrier is in a cross-section in
the trimer plane. In Mn$_{2}$N$_{3}$ the global minima are
perpendicular to the cluster plane. A high barrier is separating
the oppositely oriented states in a plane parallel to the Mn-Mn
axis, and a low barrier separates the minima in a plane
perpendicular to the Mn-Mn axis. For the Mn$_{x}$N$_{y}$ clusters,
the anisotropy is per Mn atom.
Values in parenthesis refer to the AFM configuration.}%
\label{symtab}%
\end{table}

In table \ref{bdesym}, we see that there can be a change in the
the Chern number, namely the dimensionality of the total spin
system when SOI is turned on. In general this happens when the
HOMO-LUMO gap is zero in the absence of the SOI. When SOI is
turned on, the degeneracy is lifted but in a few nearly avoided
level-crossings for particular directions of the magnetization.
Berry phase contributions coming from paths encircling these
quasi-degeneracies are responsible for the change in the Chern
number. Thus the assumption that the dimension of effective spin
the does not change when SOI is turned on is potentially a
hazardous one in the case of small TM clusters.

\subsection{Magnetic anisotropy energies}

Let us first take a look at the magnetic anisotropy energies, before
performing the variable transformation that takes us to constant curvature
space. A qualitative description of the magnetic anisotropy landscapes and the
associated anisotropy energies per atom, $|E_{max}(\hat{n})-E_{min}(\hat{n})|/N_{atoms}$,
are presented in table \ref{symtab}.

To evaluate the topological effect, we first emulate a simplified
approach to construct a quantized Hamiltonian by omitting the
variable transformation that takes us to constant curvature space.
We then take the dimension of the effective spin-system to be that
inferred by the magnetic moment in the absence of SOI (i.e. the
Chern number without SOI in table \ref{bdesym}), with the
assumption that turning it on causes no level-crossing. The
magnetic anisotropy energy is taken to be that which we obtain by
actually switching on the SOI, but should in this context be
regarded as originating in a perturbational approach. Fitting the
anisotropy energy with spherical harmonics limited by the system
dimensionality without SOI, results in the effective quantum
Hamiltonians displayed in table \ref{ham}.

\begin{table}[ptb]%
\begin{tabular}
[c]{|c||c|c|}\hline \emph{System} & \emph{Hamiltonian} &
\emph{Coefficients }[meV]\\\hline\hline Rh$_{2}$ & $\alpha
[1+\beta S_{z}^{2}+\gamma S_{z}^{4}$]$_{S=2}$ &
\multicolumn{1}{|l|}{$%
\begin{array}
[c]{l}%
\alpha=85.1,\beta=-0.630\\
\gamma=0.095
\end{array}
$}\\\hline
Ni$_{2}$ & $\alpha [1+\beta S_{z}^{2}$]$_{S=1}$ & \multicolumn{1}{|l|}{$%
\begin{array}
[c]{l}%
\alpha=-7.68, \beta=-2.0
\end{array}
$}\\\hline
Pd$_{2}$ & $\alpha [1+\beta S_{z}^{2}$]$_{S=1}$ & \multicolumn{1}{|l|}{$%
\begin{array}
[c]{l}%
\alpha=-2.44, \beta=-2.0
\end{array}
$}\\\hline
Pd$_{2}^{\ast}$ & $\alpha [1+\beta S_{z}^{2}$]$_{S=1}$ & \multicolumn{1}{|l|}{$%
\begin{array}
[c]{l}%
\alpha=66.92, \beta=-1.0
\end{array}
$}\\\hline Co$_{2}$ & $\ \ \alpha [1+\beta S_{z}^{2}+\gamma
S_{z}^{4}]_{S=2}\ \ $ &
\multicolumn{1}{|l|}{$%
\begin{array}
[c]{l}%
\alpha=45.3,\beta=-0.555\\
\gamma=0.076
\end{array}
$}\\\hline
$\text{Co}_{3}$ & $%
\begin{array}
[c]{c}%
\alpha [1+\beta S_{z}^{2}+\\
\gamma (S_{x}^{2}-S_{y}^{2})]_{S=7/2}
\end{array}
$ & \multicolumn{1}{|l|}{$%
\begin{array}
[c]{l}%
\alpha=0.491, \beta=0.554\\
\gamma=0.10
\end{array}
$}\\\hline
Co$_{3}^{\ast}$ & $\alpha [1+\beta S_{z}^{2}$]$_{S=5/2}$ & \multicolumn{1}{|l|}{$%
\begin{array}
[c]{l}%
\alpha=2.88, \beta=-0.16
\end{array}
$}\\\hline
Co$_{4}$ & $\alpha [1+\beta S_{z}^{2}$]$_{S=4}$ & \multicolumn{1}{|l|}{$%
\begin{array}
[c]{l}%
\alpha=3.94, \beta=-0.063
\end{array}
$}\\\hline
Co$_{5}$ & $\alpha [1+\beta S_{z}^{2}$]$_{S=13/2}$ & \multicolumn{1}{|l|}{$%
\begin{array}
[c]{l}%
\alpha=0.589, \beta=-0.024
\end{array}
$}\\\hline
Co$_{3}$Fe$_{2}$ & $\alpha [1+\beta S_{z}^{2}$]$_{S=13/2}$ & \multicolumn{1}{|l|}{$%
\begin{array}
[c]{l}%
\alpha=2.71, \beta=-0.024
\end{array}
$}\\\hline
Mn$_{2}$N$_{3}$ & $%
\begin{array}
[c]{c}%
\alpha [1+\beta S_{z}^{2}+\\
\gamma (S_{x}^{2}-S_{y}^{2})]_{S=5/2(1/2)}
\end{array}
 $ &
\multicolumn{1}{|l|}{$%
\begin{array}
[c]{l}%
\alpha=0.0256(0.73)\\
\beta=10.8(0)\\
\gamma=-2.91(0)
\end{array}
$}\\\hline
Mn$_{2}$N$_{3}^{\ast}$& $\begin{array}
[c]{c}%
\alpha [1+\beta S_{z}^{2}+\\
\gamma (S_{x}^{2}-S_{y}^{2})]_{S=5/2(1/2)}
\end{array}
 $ &
 \multicolumn{1}{|l|}{$%
\begin{array}
[c]{l}%
\alpha=-0.157(0.50)\\
\beta=-1.71(0)\\
\gamma=0.23(0)
\end{array}
$}\\\hline
Mn$_{3}$N$_{2} $ & $\alpha [1+\beta S_{z}^{2}$]$_{S=13/2(5/2)}$ & \multicolumn{1}{|l|}{$%
\begin{array}
[c]{l}%
\alpha=-0.358(-0.619)\\
\beta=-0.308(-0.800)
\end{array}
$}\\\hline
Mn$_{3}$N$_{2}^{\ast} $ & $\alpha [1+\beta S_{z}^{2}$]$_{S=15/2(5/2)}$ & \multicolumn{1}{|l|}{$%
\begin{array}
[c]{l}%
\alpha=-0.257(3.82)\\
\beta=-0.267(-0.160)
\end{array}
$}\\\hline
Mn$_{3}$N$_{4}$ & $\alpha [1+\beta S_{z}^{2}$]$_{S=9/2(3/2)}$ & \multicolumn{1}{|l|}{$%
\begin{array}
[c]{l}%
\alpha=-0.303(-1.82)\\
\beta=-0.444(-1.33)
\end{array}
$}\\\hline
Mn$_{3}$N$_{4}^{\ast}$ & $\alpha [1+\beta S_{z}^{2}$]$_{S=9/2(3/2)}$ & \multicolumn{1}{|l|}{$%
\begin{array}
[c]{l}%
\alpha=-0.300(1.28)\\
\beta=-0.444(-0.444)
\end{array}
$}\\\hline
\end{tabular}
\caption{The Hamiltonians obtained by quantizing in the spin
Hilbert space of the dimension given by the Chern Number without
SOI (see table \ref{bdesym}). These Hamiltonians are calculated
without performing the variable transformation to constant
curvature space, thereby neglecting the topological effect. This
approach is designed to emulate the results one would obtain by
adding the spin-orbit shifts perturbationally. The quantization
axis of the big spin is taken to be the cluster symmetry axis as
defined in table~\ref{symtab}.}
\label{ham}%
\end{table}
Generally, we find only coefficients up to second order and in
isolated cases (Co$_{2}$, Rh$_{2}$) a small fourth order
contribution in the spin operators. The higher order contribution
is an indication of level-crossings and that non-trivial
topological effects are present when SOI is switched on. If the
dimension of the Hilbert space does not change when SOI is turned
on, there is most likely no level-crossing at the HOMO. With a
large HOMO-LUMO gap, the curvature is essentially constant with
very small variations around the Chern number and there is no
topological effect that enters through the variable transform
(equations \ref{trphi} and \ref{trtheta}). In this case our
effective Hamiltonian approach will yield the same result as in
table \ref{ham}.
\begin{table}[ptb]%
\begin{tabular}
[c]{|c||c|c|}\hline
\emph{System} & $%
\begin{array}
[c]{c}%
\text{\emph{Transformed}}\\
\emph{Hamiltonian}%
\end{array}
$ & \emph{Coefficients} [meV]\\\hline\hline
Rh$_{2}$ & $%
\begin{array}
[c]{c}%
\alpha [1+\beta S_{z}^{2}+
\gamma S_{z}^{4}\\+\delta S_{z}^{6}+\varepsilon S_{z}^{8}]_{J=4}%
\end{array}
$ & \multicolumn{1}{|l|}{$%
\begin{array}
[c]{l}%
\alpha=298.79,
\beta=-2.3602\\
\gamma=0.94995,
\delta=-0.11266\\
\varepsilon=0.0038916
\end{array}
$}\\\hline Pd$_{2}^{\ast}$ & $%
\begin{array}
[c]{c}%
\alpha [1+\beta S_{z}^{2}\\+ \gamma S_{z}^{4}]_{J=2}
\end{array}
$ & \multicolumn{1}{|l|}{$%
\begin{array}
[c]{l}%
\alpha=6.19,
\beta=12.2\\
\gamma=-3.11
\end{array}
$}\\\hline
Co$_{2}$ & $%
\begin{array}
[c]{c}%
\alpha [1+\beta S_{z}^{2}+
\gamma S_{z}^{4}\\+\delta S_{z}^{6}+\varepsilon S_{z}^{8}]_{J=4}%
\end{array}
$ & \multicolumn{1}{|l|}{$%
\begin{array}
[c]{l}%
\alpha=145.30,
\beta=-2.3058\\
\gamma=0.93035,
\delta=-0.11038\\
\varepsilon=0.0038124
\end{array}
$}\\\hline
Co$_{3}^{\ast}$ & $\alpha [1+\beta S_{z}^{2}]_{J=3/2}$ & \multicolumn{1}{|l|}{$%
\begin{array}
[c]{c}%
\alpha=3.03, \beta=-0.445
\end{array}
$}\\\hline Co$_{3}$Fe$_{2}$ & $
\begin{array}
[c]{c}%
\alpha [1+\beta S_{z}^{2}\\+\gamma S_{z}^{4}]_{J=11/2}
\end{array}
$ & \multicolumn{1}{|l|}{$%
\begin{array}
[c]{l}%
\alpha=2.61,
\beta=-0.0235\\
\gamma=-0.000317
\end{array}
$}\\\hline
Mn$_{2}$N$_{3}$ & $\alpha [S^{2}]_{J=1/2}^{AFM}$ &
\multicolumn{1}{|l|}{$%
\begin{array}
[c]{l}%
\alpha=0.137
\end{array}
$}\\\hline Mn$_{2}$N$_{3}^{\ast}$ & $\alpha [S^{2}]_{J=1/2}^{AFM}$
&
\multicolumn{1}{|l|}{$%
\begin{array}
[c]{l}%
\alpha=0.163
\end{array}
$}\\\hline Mn$_{3}$N$_{2}^{*}$ & $
\begin{array}
[c]{c}%
\alpha [1+\beta S_{z}^{2}\\+\gamma S_{z}^{4}]_{J=5/2}^{AFM}
\end{array}
$ &
\multicolumn{1}{|l|}{$%
\begin{array}
[c]{l}%
\alpha=2.76,
\beta=0.365\\
\gamma=-0.0840
\end{array}
$}\\\hline
Mn$_{3}$N$_{4}^{*}$ & $%
\begin{array}
[c]{c}%
\alpha [1+\beta S_{z}^{2}]_{J=3/2}^{AFM}
\end{array}
$ &
\multicolumn{1}{|l|}{$%
\begin{array}
[c]{l}%
\alpha=1.71, \beta=-0.444
\end{array}
$}\\\hline
\end{tabular}
\caption{The Hamiltonians obtained by quantizing after performing the variable
transformation to constant curvature space. The cases in table \ref{ham} where
the transformation exhibited a very weak and negligible topological effect
have been omitted above. Only the AFM Mn$_{x}$N$_{y}$ have non-trivial curvatures.} %
\label{tham}%
\end{table}

\subsection{Effective Chern-number spin Hamiltonians}

\begin{figure}[ptb]
\resizebox{6.7cm}{!}{\includegraphics{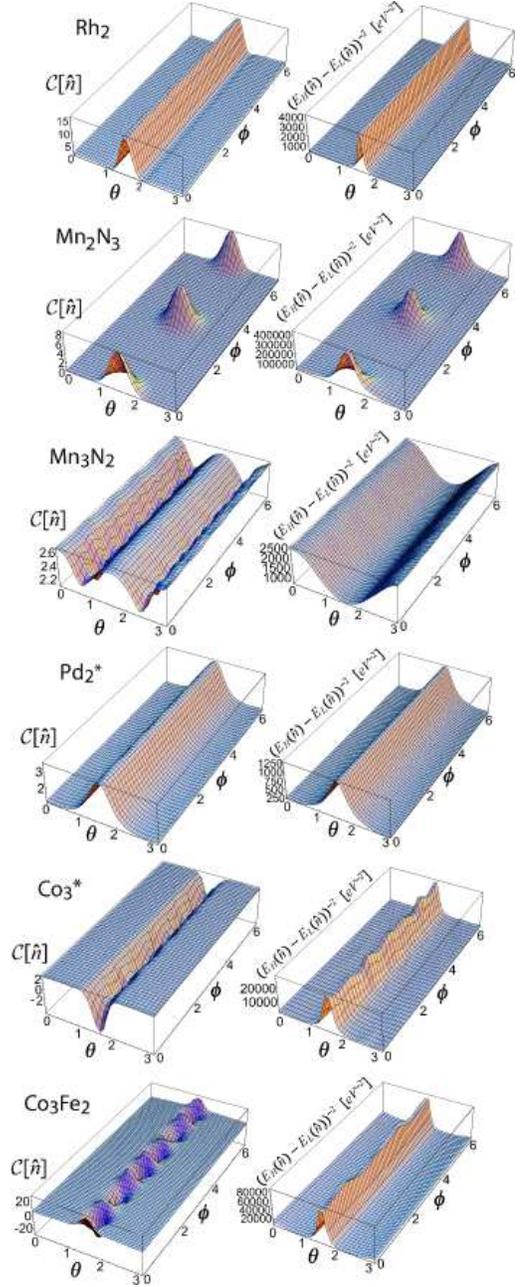}}\caption{ (Color online)
The Berry curvature and the inverse of the HOMO-LUMO gap squared on the unit
sphere for selected clusters. To the right of each curvature is the inverse of
the associated HOMO-LUMO gap squared, demonstrating the correspondence with the
perturbational expression (\ref{pertcurv}).  Rh$_{2}$ and Co$_{2}$ display very
similar curvatures that are constant at 2,
with the exception of an equatorial quasi-singular ridge that
extends up to approximately 16.
The AFM Mn$_{2}$N$_{3}$ and Mn$_{2}$N$_{3}^{\ast}$ have two
Gaussian peaks perpendicular to the cluster plane, extending from
0 to around 8. The AFM Mn$_{3}$N$_{2}^{\ast}$ sticks out as the
correspondence with the HOMO-LUMO gap is not entirely obvious. The
three lower clusters show the triggering of a quasi-degeneracy
between the HOMO and LUMO levels, by reducing the interatomic
distance in Pd$_{2}^{\ast}$, by symmetrizing the cluster geometry
Co$_{3}^{\ast}$, and by replacing the axial Co with Fe in
Co$_{3}$Fe$_{2}$.
}%
\label{curvs}%
\end{figure}

We now consider the effect on the magnetic anisotropy obtained in performing
the variable transformation to constant curvature space. Fig.~\ref{curvs} shows the total
system curvature landscapes on the unit sphere for a few clusters, as well as
the inverse of the HOMO-LUMO gap squared. This figure demonstrates that the
quasi-singular behavior of the Berry curvature landscape is governed by the
gap, as expected from the perturbational expression Eq.~(\ref{pertcurv}).

Table~\ref{tham} displays the result of quantizing the classical Hamiltonian
in a spin-space of a dimension dictated by the Chern number with SOI, using
the transformed magnetic anisotropy energy, thus incorporating the topological
effect that enters via the Berry curvature.

\begin{figure}[ptb]
\resizebox{8cm}{!}{\includegraphics{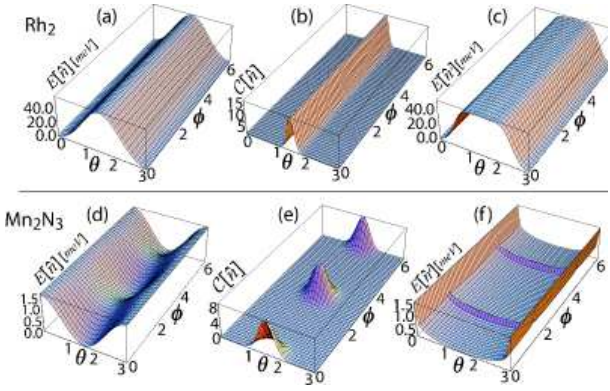}} \caption{(Color online) The
effect of the variable transform to constant curvature space for Rh$_{2}$ and
Mn$_{2}$N$_{3}$. (a) and (d) show the calculated magnetic anisotropy landscape
on the unit sphere and (c) and (f) show the transformed anisotropies where the
topological effect of the Berry curvature (panels (b) and (e)) has been
included. Rh$_{2}$ and Co$_{2}$ are topological twins and have the same
curvatures and anisotropy landscapes (anisotropies differ only in magnitude due
to the heavier Rh mass and stronger SOI). The equatorial quasi-singular ridge
in Rh$_{2}$ and Co$_{2}$ enters the transformation and causes the effective
barrier to widen significantly. The extreme Gaussian-shaped quasi-singular
peaks in the curvature of Mn$_{2}$N$_{2}$ and Mn$_{2}$N$_{2}^{\ast}$ compress
the low barriers, so that the transformed anisotropy essentially
becomes a shallow quasi-plane.}%
\label{maes3d}
\end{figure}
\begin{figure}[ptb]
\resizebox{6.5cm}{!}{\includegraphics{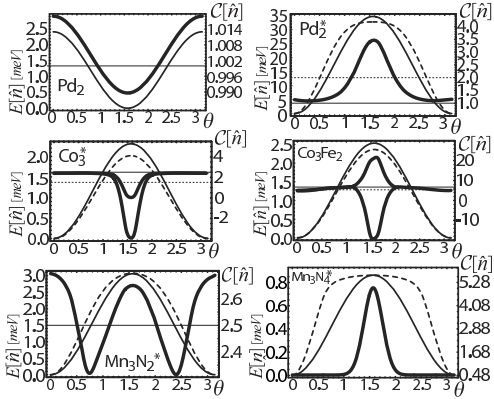}}\caption{$\theta $-dependence
of the curvature and the calculated and transformed magnetic anisotropy
landscapes for selected clusters. The thick solid line is the curvature, the
thin solid line the untransformed anisotropy energy and the dashed line the
anisotropy energy after the variable transformation to constant curvature space
has been performed. The horizontal solid lines indicate the Chern number
without SOI, and the dashed ones with SOI. Pd and Ni have the same valence
configuration and (although with different
HOMOs) are topological twins. }%
\label{maes}
\end{figure}
Fig.~\ref{maes3d} shows the magnetic anisotropy landscapes on the unit sphere
before and after the transformation to constant curvature space for Rh$_{2}$ and
Mn$_{2}$N$_{3}$, as well as the Berry curvature that enters their transform.
Fig.~\ref{maes} shows cross-sections of the unit sphere with transformed magnetic
anisotropy energies and comparative sections with the untransformed anisotropies,
as well as the Berry curvature for a few clusters.
The negative dips
at the equator for Co$_{3}^{\ast}$ and Co$_{3}$Fe$_{2}$ have been cut in the
transform in order to approximately capture the topological effect in these
systems. Note that the transform cannot really change the magnitude of the
energy values, rather, the quantization process involves the best possible fit
with the allowed spherical harmonics which will average the transformed
anisotropy energy into the subspace whose dimension is given by the Chern
number. As a result of this averaging procedure the maximal energy is affected
in accordance with the effect of the transform entering through the curvature.
In the case of Co$_{3}^{\ast}$ and Co$_{3}$Fe$_{2}$ the pull of the sub-Chern
curvature on the equator is the prevailing effect, yielding a lower effective
maximum.
Not very surprisingly, there is no
topological effect present when the HOMO-LUMO gap grows as in the case of
Pd$_{2}$, Ni$_{2}$, Co$_{3,4,5}$ and the relaxed linear Mn$_{x}$N$_{y}$
(these have therefore been omitted from
table \ref{tham}). We will now address each group of systems
and the mechanisms present in detail.

\subsection{Dimers}

\begin{figure}[ptb]
\resizebox{6.0cm}{!}{\includegraphics{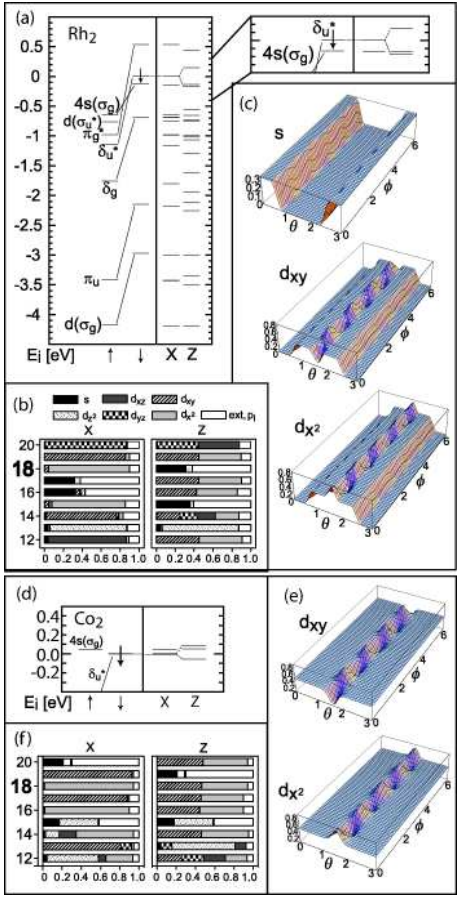}} \caption{(Color online)
Kohn-Sham orbital energies and projections of orbital character for Rh$_{2}$
and Co$_{2}$. The left panel in (a) and (d) shows the majority and
minority-spin levels in the absence of SOI and the right panel the levels in
the hard (x) and easy (z) direction with SOI. The origin of the large
anisotropy in Rh$_{2}$ and Co$_{2}$ is a first order contribution in the
spin-orbit perturbation series, resulting from the the splitting of the singly
occupied but doubly degenerate HOMO-LUMO $\delta_{u}^{\ast}$-level. The HOMO
(number 18) is connected with a dotted line and its population is marked with a
minority spin arrow. When the SOI is turned on, the quasi-degeneracies in the
hard direction split as we change the magnetization direction to the easy
(symmetry) axis. Panels (b) and (f) show the orbital mixing character as given
by the atomic site projected spherical harmonics taken over the spinors in the
indicated direction of magnetization. Panels (c) and (e) show the orbital
mixture landscapes of the HOMO on the unit sphere. The orbital mix of the HOMO
(18) and LUMO (19) show that along the easy axis these states have good
$m_{\ell}$ quantum numbers $\pm2$, allowing for a continuous electron path in a
plane perpendicular to the magnetization direction and a total lowering of the
energy. }
\label{co2spec}%
\end{figure}
Transition-metal dimers have an inherent symmetry that gives them
remarkable magnetic properties - they are rotationally invariant
around the dimer axis.
This can cause magnetic anisotropy in transition-metal dimers
to appear already at first order in a perturbational
treatment of the SOI\cite{tos_cmc_ahm_natmat2007}.
Depending on the electronic configuration and the situation at the HOMO,
this symmetry property can cause the magnetic anisotropy to
be anomalously strong.

We will discuss the dimers in terms of their Kohn-Sham
orbital energies labeled using standard molecular notation\cite{herzberg}.
Fig.~\ref{co2spec} (a) and (d) show these for Rh$_{2}$ and Co$_{2}$
(around HOMO only) without SOI for majority $(\uparrow)$ and
minority-spin $(\downarrow)$,
and with SOI and in the hard (x) and easy (z) direction.
The $s$ and $d$ valence orbitals on one atom hybridizes with orbitals
on the other atom that have the same azimuthal angular
momentum $m_{\ell}$, to form bonding and antibonding combinations.
In the absence of SOI, these are doubly degenerate and are labeled
according to their azimuthal angular momentum,
where $\sigma$ refers to $m_{\ell}=0$,
$\pi$ to $m_{\ell}=\pm 1$ and
$\delta$ to $m_{\ell}=\pm 2$. A $^{*}$
indicates an antibonding orbital and u/g labels odd/even states.
The SDFT exchange potential lowers majority-spin orbital energies relative to
those of minority-spin type.

Both Rh$_{2}$ and Co$_{2}$ have very large magnetic anisotropies
of 26 and 14 meV per atom respectively. They both exhibit an
extreme topological effect due to the presence of a
quasi-degeneracy at the HOMO level. The curvature (see
fig.~\ref{maes3d}) is constant at 2 (note that the Chern number is
2 in the absence of SOI) all over the unit sphere, except at the
equator where a quasi-singular ridge extends up to approximately
16.
Fig.~\ref{co2spec} (b) and (f) show the orbital mixing character as given by the
site projected spherical harmonics inside non-overlapping atomic spheres
taken over the spinors in the direction of the magnetization,
for the topmost eigenlevels in the hard and easy direction (as the spheres do not overlap,
some interatomic charge will be missed by the projection spheres).
Using the calculated Chern spectrum (see fig.~\ref{cspec}) in combination
with the atomic site-projected orbital character for each quasi-particle level,
we can extract the orbital and spin contribution parts to the level Chern number.
The crucial level in both Rh$_{2}$ and Co$_{2}$ is the HOMO $\delta_{u}^{\ast}$,
which is doubly degenerate but singly occupied in the absence of SOI.
From the Chern spectrum of Co$_{2}$ (fig.~\ref{cspec}) and the orbital mixing (fig.~\ref{co2spec})
we see that the HOMO (level 18) has a level Chern number of 3/2
with an orbital and spin contribution of 2 and -1/2 respectively. The orbital
angular momentum is opposing the magnetization and this level will therefore
be shifted down, whereas the LUMO-level with orbital part -2 and spin part
-1/2 will be shifted up. The shifts in the sub-levels will to a large extent
cancel each other out, but the HOMO shift remains uncompensated, causing a
very large anisotropy energy.
In Rh$_{2}$ the spin-orbit shift is so strong that the $\delta_{u}^{\ast}$
descends below the minority-spin 4s($\sigma_{g}$), which is much higher
in energy for Rh$_{2}$ than for Co$_{2}$ - consistent with the larger dimer
separation.

Up to double-counting corrections, the total energy is simply the sum of the
spin-orientation-dependent shifts in occupied orbital energies\cite{ac_cmc_ahm2002}.
In the hard direction spin-orbit shifts are small compared with the scale
of the $d$-electron spin-orbit interaction ($H_{SO}=\xi_{d}\vec{s}\cdot\vec{L}$,
with $\xi_{d}^{Co}\sim 85$ meV and $\xi_{d}^{Rh}\sim 140$ meV).
With the spins polarized in the hard (x) direction the expectation value of $L_{x}$
over the unperturbed states is zero, and any shifts originate in higher order
perturbation terms (at most $\xi_{d}^2/W_{d}\sim0.1\xi_{d}$, where $W_{d}$ is
the typical $d$-orbital bonding energy).
However, when the spins are polarized along the easy (z)
direction, i.e. the dimer axis, the first order contribution is non-vanishing
and is given by $\pm \xi_{d} \langle L_{z} \rangle = \pm m_{\ell} \xi_{d} /2$.
For Rh$_{2}$ and Co$_{2}$ the first order contribution is larger than the obtained
anisotropy energy and the expected $|\cos(\theta)|$ appearance is not
observed.  This is because the cusp at the degeneracy point ($\theta=\pi/2$)
is always rounded out by higher order terms, and the anisotropy energy
is an even analytic function of  $\cos(\theta)$.
The degree of rounding is a complex many-body issue,
which in our DFT calculations is affected by the smearing
of occupation numbers at the HOMO. To go beyond our calculations,
one would have to employ a true many-body framework, where the
absence of the otherwise necessary smearing parameter would
imply even larger anisotropies.

From fig.~\ref{co2spec} it is clear that the total Chern number of
4 originates in equal parts from spin and orbital angular
momentum. Without SOI there is no angular momentum contribution
and in this case the total Chern number is 2 coming from the spins
in the $d$-shells. With SOI the quasi-degeneracy is lifted along
the easy (symmetry) axis and the total energy is lowered via the
mixing of $d_{xy}$ and $d_{x^{2}-y^{2}}$ into axial eigenstates
with $m_{\ell}=\pm 2$ maximal splitting and a continuous electron
path in a plane perpendicular to the easy axis magnetization
direction. At the equator, the $\delta_{u}^{\ast}$ levels approach
each other and the curvature becomes quasi-singular. Here the
$d_{xy}$ and $d_{x^{2}-y^{2}}$ characters separate and we find
that the HOMO in the $x$-direction ($\phi=0$) is of pure
$d_{x^{2}-y^{2}}$ character (see fig.~\ref{co2spec} (c) and (e)).

The effect of the
equatorial curvature ridge in our variable transformation of the effective
big-spin Hamiltonian, is to alter the local metric (\ref{metric}) and repel
grid-points from the equator in order to increase the subtended curvature areas
in this region.
This leads to a transformed magnetic anisotropy landscape that has
a much wider barrier (see fig.~\ref{maes3d}). Such an
enormous increase in effective barrier width should definitely lead to
experimentally observable effects in terms of spin-lifetimes and tunneling probability.

The Chern number is 4, which tells us that we may use spherical harmonics of
up to order $L=8$ in our quantization process. It is interesting to note that
we actually need all of these to accurately describe the transformed magnetic
anisotropy landscape. Omitting the smallest 8:th order coefficient leads to
the formation of a non-physical shallow local minimum along the equator.

\begin{figure}[ptb]
\resizebox{7.0cm}{!}{\includegraphics{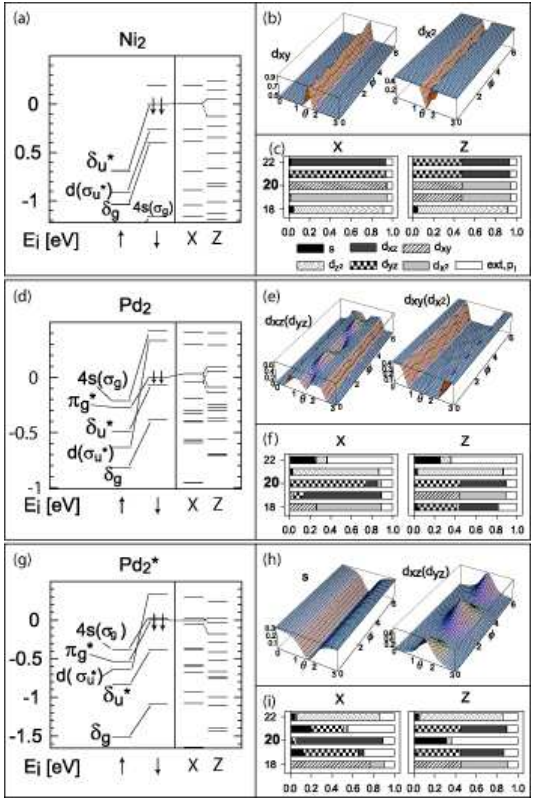}} \caption{(Color online)
Eigenlevel structure around the HOMO and atomic site projected orbital
character.  for the easy (x for Ni$_{2}$ and Pd$_{2}$, z for Pd$_{2}^{\ast}$)
and the hard (z for Ni$_{2}$ and Pd$_{2}$, x for Pd$_{2}^{\ast} $) direction,
as well as over the unit sphere for the HOMO. The orbitals in parenthesis are
the same as the ones displayed, only with an equatorial
$\phi$-dependence that is in relative anti-phase.}%
\label{nipdcombo}%
\end{figure}

Paralleling the close relationship between Co and Rh, Pd and Ni have similar
valence electron configurations - the 4d elements behave like their 3d
counterparts. In the absence of SOI, both Co$_{2}$ and Rh$_{2}$
have a molecular groundstate $\Delta(S=2)_{g}$ and a Chern number $J=4$ with SOI.
Pd$_{2}$ and Ni$_{2}$  both have a Chern number of 1 (with
and without SOI) with a $\Sigma(S=1)_{g}$ ground state.
Their anisotropy landscapes are quasi-easy planes perpendicular to the dimer line,
although the anisotropy energies are small -
balancing rather delicately between easy
axis and easy plane. The curvature is essentially constant and equal to the
Chern number, wherefore no topological effect enters via the variable
transform (see fig.~\ref{maes}). The combined $s$-shell in the dimer, demotes two atomic
$s$-electrons to the $d$-shell, leaving only two $d$-holes in the dimer. The
net $d$-spin contribution yields the Chern number 1, with no orbital
contribution. Pd$_{2}$ has a much larger dimer separation than Ni$_{2}$, which
results in a different ordering of the levels.
Fig.~\ref{nipdcombo} (d) reveals that the $\pi_{g}%
^{\ast}$-level is lower in energy than the $4s(\sigma_{g})$ in
Pd$_{2}$, resulting in a $\pi_{g}^{\ast}$ HOMO, whereas Ni$_{2}$
retains the $\delta _{u}^{\ast}$ (fig.~\ref{nipdcombo} (a)). In
both cases, the doubly degenerate HOMOs are doubly occupied. The
antipodal shifts now cancel out just like for most of the
sub-levels. This highlights the situation for which high
anisotropy and strong topological effects are present: when there
is a \emph{doubly degenerate} and \emph{singly occupied} HOMO in
the absence of SOI. The quasi-degenerate occupied levels in
Pd$_{2}$ and Ni$_{2}$ both have a highly singular level curvature,
but these also cancel out, resulting in a trivial total curvature
with very small variations around the Chern number. It is clear
that the case of a singly occupied, doubly degenerate HOMO is the
most natural candidate for a system in which there is a powerful
topological effect. However, Pd$_{2}^{\ast}$ points to another
possibility. In reducing the dimer distance - the $s$-level drops
in energy, until it actually intersects the HOMO-LUMO gap. In
doing so, the strong topological effect (see fig.~\ref{curvs}) is
triggered due to the mixing with the $\pi_{g}^{\ast}$ doublet (see
fig.~\ref{nipdcombo} (g)). The non-trivial curvature enters the
transform and causes the effective barrier for Pd$_{2}^{\ast}$ to
widen and the effective Hamiltonian acquires a fourth order
contribution (see table \ref{tham}). Note the inverse relation
between the $s$-character of the HOMO (fig.~\ref{nipdcombo} (h))
and the curvature (fig.~\ref{curvs}), i.e. the curvature barrier
grows as the $s$-character diminishes and the
$d_{xz}(d_{yz})$-character increases.

The situation observed in Pd$_{2}^{\ast}$,
where an accidental degeneracy triggers a topological effect, by
intermixing with the HOMO-LUMO doublet, can also occur in relaxed systems. For
example, in C$_{2}$ \cite{pederson_prb91:Pfrac_occ}, there is the exact same
mechanism as in Pd$_{2}^{\ast}$ present at the HOMO, where an $s$-level causes
an accidental degeneracy with a $\pi$ HOMO, for a wide range of dimer distances.

\subsection{Co$_{3}$ and Co$_{3}^{\ast}$}

\begin{figure}[ptb]
\resizebox{8cm}{!}{\includegraphics{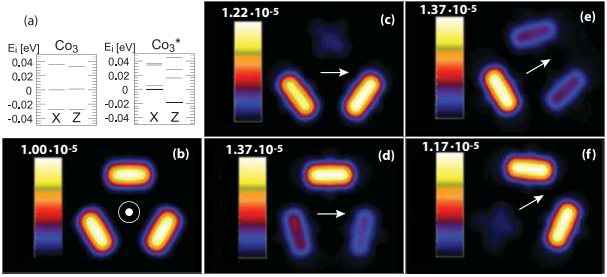}} \caption{(Color online)
Voxel projections of the Co$_{3}^{\ast}$ HOMO and LUMO charge densities for
different directions of the magnetization. (a) shows the energies of the HOMO
and the LUMO for the isosceles and the equilateral trimer in the hard and easy
direction. (b) shows the HOMO charge density for the equilateral trimer in a
top view using maximal intensity projection. Here, a small amount of $d$-charge
forms axial states and the HOMO-LUMO splitting is maximal, lowering the total
energy. (c) and (e) show the HOMO charge density when the magnetization is in
the hard plane along the equator. Here, the $d$-character separates, occupying
the same sites as the $p$-orbitals. (d) and (f) show the corresponding LUMO
charge densities which occupy increasingly disjoint regions of the cluster as
the gap closes in.
}%
\label{co3vox}%
\end{figure}
Just like the symmetrized Co$_{4,5}$, the equilateral trimer Co$_{3}$ has a
highly degenerate level structure and are therefore unstable against
Jahn-Teller distortions (see fig.~\ref{positions}). Fig.~\ref{co3vox} (a) shows
the levels in close vicinity of the HOMO in the hard (x) and easy (z)
direction for Co$_{3}^{\ast}$ and the corresponding directions for the
Jahn-Teller-distorted Co$_{3}$, where the hard and easy direction is
interchanged relative the equilateral structure.

For the Co$_{3}$ the orbital mixing changes very little between the two
directions and there is a large HOMO-LUMO gap of 0.03-0.04 eV. By contrast,
the Co$_{3}^{\ast}$ has a quasi-degeneracy in the hard direction of approximately 4
meV that is lifted to around 35 meV in the easy direction.
 The quasi-degeneracy that is
lifted by the Jahn-Teller distortion, induces large variations from the Chern
number in the curvature (see fig.~\ref{curvs} and \ref{maes}). In
fig.~\ref{cspec} we see that the level Chern number for the HOMO is -3/2,
meaning that the HOMO level curvature that goes quasi-singular has negative
curvature. Although this negative curvature is shifted by the accumulated
Chern number coming from the occupied sub-levels (level 26 has an accumulated
Chern number of 3 in fig.~\ref{cspec}), negative curvature residues remain, implying
that we can only approximately describe the topological impact by omitting the
negative dips in the transform.

Fig.~\ref{co3vox} shows voxel projections of the HOMO and LUMO
charge densities for different directions of the magnetization.
When the magnetization is in the easy (z) direction, perpendicular
to the trimer plane, each atom forms a different combination of
on-site $d$- and $p$-character, symmetrically around the z-axis
(see fig.~\ref{co3vox} (b)).  Here, a small mixture of
approximately 10\% $d_{xy}$ and $d_{x^{2}-y^{2}}$ in the whole
system is responsible for the lowering of the HOMO energy. On the
equator (see fig.~\ref{co3vox}) (b)-(f)), these separate,
different atoms are occupied by the HOMO charge
 and the HOMO-LUMO become quasi-degenerate. The splitting increases slightly
 along the equator as the magnetization changes to $\phi=\frac{\pi}{6}$ (see fig.~\ref{co3vox} (e)),
 enough to produce the observed oscillations in the curvature. In the hard direction,
 the LUMO density occupies increasingly disjoint regions of the cluster, as the HOMO-LUMO
 gap closes in.  The occupied configuration where the atoms are unequally occupied is higher
 in energy than the symmetric configuration in along the $z$-axis,
resulting in a bistable configuration with a total magnetic anisotropy energy of 3 meV.

\subsection{Co$_{5}$ and Co$_{3}$Fe$_{2}$}

\begin{figure}[ptb]
\resizebox{7.5cm}{!}{\includegraphics{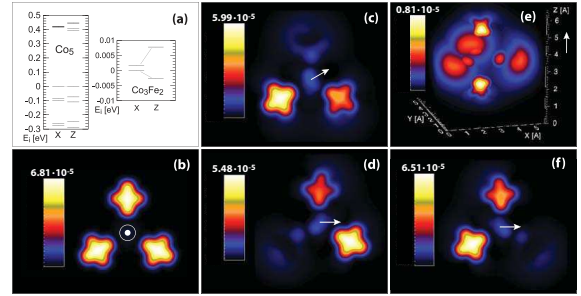}} \caption{(Color online)
The level structure around the HOMO for Co$_{5}$ and Co$_{3}$Fe$_{2}$ and voxel
projections of the Co$_{3}$Fe$_{2}$ HOMO charge density for different
magnetization directions. (a) shows that the substitution of the axial Co atoms
in Co$_{5}$ for Fe closes the HOMO-LUMO gap from approximately 0.4 to
0.002-0.01 eV, causing an extreme quasi-degeneracy and impact on the Berry
curvature. (b) shows a maximal intensity projection of the HOMO charge density
when the magnetization direction is parallel to the bi-pyramidal easy-axis (top
view). (c) and (d) show the HOMO densities with the magnetization in hard
plane. The variations in $d$-character are responsible for the equatorial
oscillations in the curvature. In (e) and (f) the LUMO charge densities when
the magnetization is in the $z$- and $x$-direction are shown. }
\label{co3fe2vox}%
\end{figure}
Diagrams of the level structure around the HOMO for
Co$_{3}$Fe$_{2}$ and Co$_{5}$ are shown in fig.~\ref{co3fe2vox}
(a). When the axial Co atoms in Co$_{5}$ are substituted for Fe,
the character of the HOMO changes from $p$- to $d$-like. In
addition, the HOMO-LUMO gap decreases from approximately 0.4 eV in
Co$_{5}$ to 2-10 meV in Co$_{3}$Fe$_{2}$. This dense level
structure in Co$_{3}$Fe$_{2}$ causes a level-crossing when SOI is
turned on, changing the Chern number from 13/2 to 11/2. In
Co$_{5}$ however, there is no level-crossing, and the Chern number
does not change with the inclusion of SOI.

Fig.~\ref{positions} shows that the substitutional Fe atoms compress the
cluster along the bi-pyramidal axis and the center triangle increases in
size. This points to the presence of an attractive force between the Fe atoms
that is counteracted by the Co triangle, causing a more symmetric system with
an accidental degeneracy at the HOMO.
The anisotropy/atom (see table \ref{symtab}) increases by a factor five with
the substitution - largely due to the uncompensated HOMO-downshift. Pederson
et.~Al \cite{mark_apl02:_cofe} report a matching anisotropy/atom of 0.1 meV and
moment of 13 $\mu_{B}$ for Co$_{5}$ with an increase in anisotropy/atom with a
factor of approximately 5 in Co$_{3}$Fe$_{2}$.
The moment of 13 $\mu_{B}$ 
matches our non spin-orbit Chern number, but this changes as SOI is included
(see table \ref{bdesym}). Since the large fluctuations in the curvature at the
equator (see fig.~\ref{curvs})
cancel out to a large extent in the $\theta$-variable transform (\ref{trtheta}%
), and since there is no $\phi$-dependence in the anisotropy landscape, the prevailing
topological effect can be captured in the transform by cutting the negative
dips (see fig.~\ref{maes}).

Fig.~\ref{co3fe2vox} (b)-(d) show maximal intensity projections of
the HOMO charge density along the easy (z) axis and along two
symmetry directions along the hard plane (the Co$_{3}$-plane).
These images reveal that there is not much HOMO charge on the Fe
atoms and that the dynamics of the HOMO is governed by the
Co$_{3}$-triangle. In Fig.~\ref{co3fe2vox} (b) we see that the Co
do not really form complete axial states, rather the high
anisotropy comes from the fact that the states corresponding to
the hard plane (Fig.~\ref{co3fe2vox} (c) and (d)) are relatively
higher in energy. The variation of the $d$-orbitals on the
different Co when the magnetization changes in the hard plane is
responsible for the large equatorial oscillations. In
fig.~\ref{co3fe2vox} (e) and (f) the LUMO charge densities are
shown for a magnetization in the $z$ and $x$-direction
respectively. The $z$ LUMO charge density has maximum on the Fe
atoms, and is much higher in energy than the symmetric Co triangle
in the HOMO.

\subsection{Mn$_{x}$N$_{y}$}

\begin{figure}[ptb]
\resizebox{7cm}{!}{\includegraphics{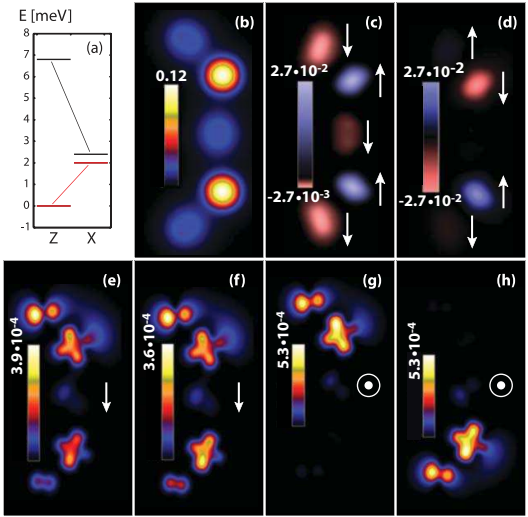}} \caption{(Color online)
Energy levels and charge densities for the HOMO and the LUMO in
Mn$_{2}$N$_{3}$. (a) shows the HOMO and LUMO energies in the hard (z) and easy
(x) direction. In (a), (b) and (c), the total charge density, the magnetization
density in the FM variation and in the AFM variation are shown, respectively.
(e) and (f) show the charge density of the HOMO and the LUMO in the
$z$-direction
and (g) and (h) in the $x$-direction.}%
\label{mn2n3vox}%
\end{figure}

Finally, we turn our attention to the hybrid  Mn$_{x}$N$_{y}$ clusters, partly
motivated by the work of Hirjibehedin et. Al. \cite{Hirjibehedin_sci06:_MnCuN}.
In order to properly simulate their system, it is necessary to include some layers of Cu, which,
although beyond the scope of this article, certainly warrants further investigation.
As the interesting topological effects occur only for the AFM configurations,
we shall focus primarily on those.

We start with the Mn$_{2}$N$_{3}$-clusters. In Mn$_{2}$N$_{3}^{\ast}$ the
atoms have been locked in a configuration that corresponds to their
positions as they are embedded in the covalent network of the surface. This cluster does not differ much from
the Mn$_{2}$N$_{3}$, where the Mn have been frozen at a distance of 3.6 A and the
N where allowed to relax starting from off Mn-axis positions. The angle of the endpoint
N is some what steeper and the distance to the Mn longer because of the presence of the surface Cu atoms.

Fig.~{\ref{mn2n3vox}} (a) show how the HOMO-LUMO levels vary
between the easy (x) and the hard (z) direction, for
Mn$_{2}$N$_{3}$. In Mn$_{2}$N$_{3}$ the gap varies between 7 and
0.5 meV between the hard and easy direction, where as the
Mn$_{2}$N$_{3}^{\ast}$ has the same qualitative shifts but with a
smaller variation from 3.5 to 1 meV. In Fig.~{\ref{mn2n3vox}} (b)
the total charge density is shown, where the interspacing N atoms
hold much less valence charge. (c) and (d) show the magnetization
density in the Mn-axis direction, for the FM and the AFM variation
respectively. The fact hat the N atoms magnetize oppositely the Mn
atoms in the FM configuration, is an indication that the N
stabilize the AFM configuration. The FM and the AFM configurations
are rather close in energy (see table \ref{symtab}) although the
AFM is lower in energy by the order of ten meV. As noted in by Rao
et Al.~\cite{rao_prl02:_MnN}, the presence of N in small Mn
clusters, dramatically increases the binding energy and generally
enhances the Mn magnetic moments, highlighting the role of the N
as mediators of the AFM coupling. We find a Chern number of 1/2 in
the AFM and 5/2 in the FM variation, which indicates that the N
contribute to the effective cluster spin dimension. This
configuration could change by the inclusion of surrounding surface
Cu atoms. In the case of the dimer, a Heisenberg Hamiltonian
fitted to experimental data does not lead to distinguishable
results between different effective atomic Mn spins.

Fig.~{\ref{mn2n3vox}} (e) and (g) show the HOMO and (f) and (h)
the LUMO charge density, in the hard (z) and easy (x) direction
respectively for the Mn$_{2}$N$_{3}$ (Mn$_{2}$N$_{3}^{\ast}$
follows a similar pattern, although less pronounced). The $p$-like
charge on the N connects with the $d$-like Mn charge in one end of
the cluster. The charge configurations in fig.~{\ref{mn2n3vox}}
(e) and (f) results in a higher energy for the LUMO (level Chern
number $-1/2$) and a lower energy for the HOMO (level Chern number
$1/2$). As the magnetization is changed to the $x$-direction, the
quasi-degenerate HOMO and LUMO occupy completely disjoint regions
of the cluster. The energy difference between these symmetric
mirror configurations is very small, resulting in severe
quasi-degeneracies. These occur perpendicularly to the cluster
plane and appear in the form of Gaussian shaped
quasi-singularities in the curvature (see fig.~\ref{maes3d}).
Their effect in the transform is to compress the low barriers to a
thin line. After quantization this yields an effective quasi-plane
Hamiltonian with reduced anisotropy (see tab.~\ref{tham}). The
HOMO shift is not large enough to turn the $z$-direction into the
easy direction, it does however reduce the anisotropy somewhat. In
this case the MAE originates in the lower levels, whereas the its
non-trivial topology is coming from the HOMO-LUMO level. The
quasi-degeneracy that appears is the most severe one that we have
encountered (see fig.~\ref{curvs}). In view of this, here we
expect that our single spin procedure to break down. Nevertheless,
the transform gives a good indication of what is going on - the
anisotropy is basically being erased by the topology of the
system. The absence of spin-flip excitations in the conductance
spectrum\cite{Hirjibehedin_sci06:_MnCuN} could very well be a
manifestation of this powerful topological effect, even when the
Chern number is different from zero.
\begin{figure}[ptb]
\resizebox{6.0cm}{!}{\includegraphics{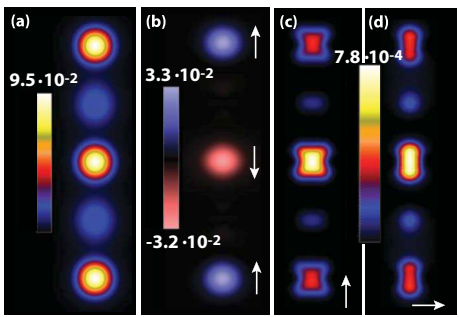}} \caption{(Color online)
Densities for Mn$_{3}$N$_{2}^{\ast}$. (a) shows the total charge density and
(b) the magnetization density in the easy (z) direction. (c) and (d) show the
HOMO charge densities in the easy and hard direction, with perspective views in
(e) and (g). (f) and (h) show the LUMO for the relaxed
cluster which is 0.3 eV above the HOMO. }%
\label{mn3n2vox}%
\end{figure}

We now address the Mn$_{3}$N$_{2}^{\ast}$. The AFM variation with Chern number
5/2 has a much lower energy than the FM variation with Chern number 15/2. This
is consistent with representing the Mn with atomic spins of 5/2 and the
anisotropy is of the order of what one would expect from the conduction
spectrum. Mn$_{3}$N$_{2}^{\ast}$ sticks out as there is no obvious correlation
between the inverse squared of the HOMO-LUMO gap and the Berry curvature (see
fig.~\ref{curvs}). The gap size does not vary that much between hard and easy
directions and is quite large, between 20-40 meV. In addition, only the HOMO
moves as the magnetization direction is changed. The HOMO up-shift is in fact
compensated by the sub-HOMO level, whereas the LUMO does not move at all. This
is the reason for the small anisotropy the odd appearance of the curvature.
Fig.~{\ref{mn3n2vox}} (a) shows the charge density and (b) the magnetization
density in the easy-direction (along the cluster axis), revealing no magnetic
charge on the N atoms. (c) and (d) show the HOMO charge density in the easy-
and hard-direction respectively. Varying $\theta$ from 0 to $\pi$, we see that
the Mn $p$- and $d$-character traces out a contour like the one observed in the
curvature. The curvature causes a slight widening of the blocking barrier (see
fig.~\ref{maes}) with a small fourth order contribution in the effective
Hamiltonian (see tab.~\ref{tham}) When Mn$_{3}$N$_{2}^{\ast}$ is allowed to
relax into Mn$_{3}$N$_{2}$ (see fig.~\ref{positions}), the N position
themselves relatively closer to the endpoint Mn, and the HOMO now shows no
charge on the center Mn. The endpoint Mn and the N now form orbital
configurations where the LUMOs are approximately 0.3 eV above the HOMO,
resulting in a trivial and constant curvature. This highlights the crucial role
of the epitaxial registration and the symmetry enforced by the surface. Adding
two N atoms on the endpoint to create Mn$_{3}$N$_{4}^{\ast}$ we find that the
Chern numbers change to 9/2 for the FM and 3/2 for the AFM variation,
consistent with instead representing the Mn with spin 3/2. Clearly the extra N
alter the Mn electronic configuration, pointing to a sensitive dependence on
the endpoint N positions. The FM variation yields a quasi-easy plane with a
lower global minimum, and only in the AFM variation of Mn$_{3}$N$_{4}^{\ast}$
do we find an easy axis parallel to the cluster axis. Relaxing the atoms on a
line again returns the AFM to the quasi-easy plane configuration.
The Mn$_{3}$N$_{4}^{\ast}$ easy HOMO charge exhibits
$p_{z}$-character on the endpoint N, but the hard direction HOMO
charge has a dominant $d_{xy}$-character on the center Mn that is
17 meV higher in energy. This level is very close in energy to the
hard direction LUMO, with a $d_{x^{2}-y^{2}}$-character - just a
rotated version of the hard HOMO. The non-trivial Berry curvature
manifests itself precisely because of this rotational symmetry.
The large shift in the HOMO of 17 meV is enough to tip the scales
and change the MAE from quasi-easy plane in Mn$_{3}$N$_{4}$ to an
easy axis in Mn$_{3}$N$_{4}^{\ast}$. The quasi-degeneracy in the
hard plane causes a widening of the blocking barrier (see
fig.~\ref{maes}). Quantizing with $J=3/2$ entails an averaging
into the allowed spin-space, such that the widened barrier is
represented by a greater effective barrier height.

\section{Alternative approach at a degeneracy point}
\label{alt_app}

The extensive numerical studies presented above show that close to
a HOMO-LUMO degeneracy we face the most interesting and {\it
difficult} situation, where our giant-spin formalism is pushed to
its limit: on one hand the (partly avoided) level-crossing signals
that Berry curvature effects are important and large Berry
curvature fluctuations modify the effective energy anisotropy
lansdcape. On the other hand, when the HOMO-LUMO gap becomes too
small, the ensuing singularities in the Berry curvature render the
implementation of our formalism problematic, particularly when the
total curvature becomes negative and the variable transformations
given in Eqs.~\ref{trtheta}, \ref{trphi} are not valid. In this
case the description of the low-energy dynamics in terms of one
single "giant-spin" breaks down, and it is necessary to
reintroduce explicitly the relevant electronic degrees of freedom
responsible for the non-adiabaticity. We sketch here an
alternative treatment for the spin dynamics that should remain
valid when a true electronic degeneracy is present. Our idea is
based on an analogy of our problem with the Jahn-Teller effect
(JT) in the vibronic studies of polyatomic molecules. (For a
review of the JT effect see Ref.~\onlinecite{bersuker_book06}.) In
the static JT effect, an electronic level degeneracy, usually
present for a very symmetric molecular configuration, is lifted by
a distortion of the molecule to a lower-symmetry configuration,
accomplished via a coupling of the electronic variables with the
nuclear degrees of freedom. The molecular symmetry, reduced in the
static JT effect, is restored by coherent tunneling between
equivalent distortions, a phenomenon known as dynamical JT (DJT)
effect.

The coupled dynamics of the (fast) electronic degrees of freedom and the (slow)
nuclear motion is usually investigated in two steps within the adiabatic
Born-Oppenheimer (BO) approximation.
Here the electronic problem is first solved for
a fixed arbitrary configuration of the nuclear variables, assumed in the
first instance to be classical. The resulting
electronic energies act as potential energy functions
for the nuclear dynamics problem,
which is then treated quantum mechanical. It turns out that near an electronic
degeneracy giving rise to the JT phenomena, non-adiabatic corrections
to this procedure in the
form of Berry phases must be taken into account. The important role
of such non-adiabatic corrections was already recognized in early studies of
electron-lattice interactions in molecules
(see Ref.~\onlinecite{wilzek_book89}
for a review of this work).
However, it was after Berry's seminal work\cite{originalberry_prs84}
that these corrections were recognized to
be examples of the ubiquitous quantum geometric phases often arising in
semiclassical treatments of complex quantum systems containing
slow and fast degrees of
freedom.
The study of DJT effects in molecular systems remains a topic of great
interest, which presently includes investigations of novel nanostructures
carried out with the help of powerful electronic structure calculations.
In this context it has recently been pointed out\cite{bersuker_prl06}
that electronic structure
calculations generating global minimum configurations and associated
vibrational frequencies might be faulty if Berry phase effects, when present,
are neglected. In general Berry phase terms affect profoundly the
quantum spectrum of the vibronic dynamics.

In order to construct our analogy with the DJT effect, we first
review the classical case of the JT $E \otimes e$ problem (a
doubly-degenerate electronic $E$ term interacting with
doubly-degenerate $e$ vibrational
mode)\cite{zwanzinger_87,koizumi_bersuker_prl99,qian_niu_book}.
The Hamiltonian for the two electronic states $\{|a\rangle,
|b\rangle \} $ and the doubly degenerate normal vibrational
coordinates $Q_1 = \rho \cos \theta$,  $Q_2 = \rho \sin \theta$ is
$H = H_0 + H_1,$
where $H_0 = \sum_{i = a,b} h_{\rm 2dho}\, |i\rangle \langle i|$,
with $h_{\rm 2dho} (\rho, \theta)$ being the Hamiltonian of a
two-dimensional isotropic harmonic oscillator describing the free
normal modes. $H_1 = \Big ( k \rho e ^{-i\theta} + \frac{1}{2} g
\rho^2 e ^{i2\theta} \Big) |i\rangle \langle i| + {\rm H.c.}$ is
the vibronic interaction Hamiltonian. The coordinates $Q_1 = Q_2
=0 \leftrightarrow \rho= 0$ represent the symmetric (undistorted)
configuration of the molecule. The adiabatic BO potential surfaces
$E_{\pm}$ are obtained by diagonalizing the electronic part of the
$H$; that is, the nuclear kinetic energy is set to zero and the
vibrational coordinates are treated as classical variables. For
the linear $E \otimes e$ JT problem, for which $g=0$, $E_{\pm} =
\frac{1}{2}\rho^2 \pm k\rho$, and the corresponding adiabatic
electronic wave functions are $|\pm (\theta) \rangle =
(e^{-i\theta/2}|a\rangle \pm e^{i\theta/2}|b\rangle)\sqrt{2}$. The
lowest potential energy surface has a continuous minimum at
$\rho_0= k$, which gives rise to the JT effect. The electronic
wave functions $|\pm (\theta) \rangle$ are double-valued when
transported adiabatically around the degeneracy in the electronic
spectrum ($E_{+} =E_{-}$) occurring at $\rho =0$. The sign change
in $|\pm (\theta) \rangle$ upon encircling the origin is a
manifestation of the topological Berry phase arising from the
conical intersection of the adiabatic potential surface. The sign
change has dramatic effect on the quantum vibration spectrum. The
simplest way to see this is to consider the large $k$ regime,
where the two potential surfaces are well separated. We can
therefore quantize the nuclear motion by restricting to the lowest
energy surface $E_-$. The relevant potential energy to be added to
$h_{\rm 2dho}$ is therefore $V = E_{-} + H_{\rm BH}$, where
$H_{\rm BH}\equiv \frac{1}{2}\sum_{i=1,2} {\partial\langle
-(\theta) |\over \partial Q_i}
{\partial|-(\theta)\rangle\over\partial  Q_i }$ is the Born-Huang
term\cite{zwanzinger_87,koizumi_bersuker_prl99,qian_niu_book},
\footnote{This term is generated when the kinetic energy of the
slow (nuclear) variables is written in terms of adiabatic basis
$|\pm (\theta)\rangle$. See Shankar's book on quantum mechanics
for an elementary derivation.}.

Since the total vibronic wave function
$\Psi(\rho,\theta)_{-} = |\rho,\theta\rangle_{-}\, \otimes |- (\theta) \rangle$
must be single-valued,
the nuclear part  $|\rho,\theta\rangle_{-}$
must change sign when going around the origin to compensate
the double-valuedness of $|-(\theta)\rangle $.
This boundary condition causes the
spectrum of $h_{\rm 2dho}$ to have half-odd integral vibronic angular
momentum quantum numbers. Alternatively,
the sign change of $|-\rangle (\theta)$
can be absorbed by re-defining its phase. This phase change introduces a
fictitious magnetic vector potential in the kinetic energy of
the nuclear motion,
known as Mead-Berry vector potential or Mead-Berry connection.
whose curl is the Berry curvature. The presence of this fictitious magnetic
field causes the same half-integer quantum numbers for the vibronic spectrum.

Quadratic and higher coupling terms in $H_1$ are in general important and
give rise to more complex adiabatic
potential energy surfaces. In particular, other conical intersections
are possible, apart from the one at $\rho= 0$. Typically in the $E \otimes e$
problem three other degeneracies are possible, each of them generating
a Berry-Mead vector potential. In this case the overall behavior
of the electronic wave
functions and the resulting vibronic spectrum depend crucially on the
dominant tunneling paths connecting potential energy minima,
{\it around these degeneracies}: if paths encircling
all four degeneracies are possible, the overall Berry phase is $4\pi$, that
is, the wave functions are single-valued and the vibronic quantum numbers are
integers.

In our problem with magnetic clusters, the slow nuclear motion is replaced
by the collective spin degree of freedom of the cluster $\vec S = S \hat n$.
This is couple to two or more electronic levels, including the HL levels,
which are degenerate for given spin orientations.
It is important to clarify that in this case we are not interested
in the nuclear dynamics: the molecule configuration is assumed to be fixed.
The Hamiltonian for the collective spin and the HL electronic levels
$\{ |H\rangle , |L\rangle\} $ is
\begin{equation}
H = H_0 + H_1\; ,
\label{ham_SHL}
\end{equation}
where
\begin{equation}
H_0 = \sum_{i = H,L}
\Big({\vec S^2 \over 2I} + U(\vec S)\Big )|i\rangle \langle i|\; ,
\end{equation}
with $V(\vec S)$ being the anisotropy energy functional without the
contribution of the HL levels. Here we take $S$ to be the total spin
moment of the cluster when SO coupling is switched off.
The term $H_1$ describes the coupling between the collective spin
and the HL electronic levels, and is written as
\begin{equation}
H_1 = \sum_{i = H,L}V_{i}(\vec S) |i\rangle \langle i|
+ \Big( V_{HL}(\vec S) |H\rangle \langle L| + H.C.\Big)\;.
\end{equation}
The functions $V_{L}$, $V_{H}$, and $V_{HL} = V_{LH}^*$ have to be chosen so
that the eigenvalues $E_{\pm}(\vec S)$ of $H_1$ reproduce the HL energy
landscapes, with the associated intersections as a function of the orientation
of $\vec S$. The eigenvalue and eigenvectors of Eq.~\ref{ham_SHL} can be
obtained numerically by expanding in the basis set $|\pm \vec S\rangle\otimes
|m, S$, where $|m, S$ are eigenvectors of $S^2$ and $S^z$. However, since we
want to single out the effect of the Berry phase on the slow variable dynamics,
it is useful to think in terms of the adiabatic semiclassical approach and use
an approximate spin Hamiltonian that includes the lowest energy surface surface
$E_{-}(\vec S)$ only. Like for the DJT problem, a conical intersection will
generate a Berry phase in the electronic wavefunction, which in turn will
affect the quantum spectrum of the collective spin variable $\vec S$. In the
presence of several conical intersections, the overall effect of the Berry
phase on the electronic wave function (and therefore on the dynamics of the
collective spin) will depend on the character of the dominant tunneling paths
connecting nearby energy minima through different saddle points around the
degeneracies. As an example we can consider the case of the Co trimer. When the
cluster is fixed in the static JT distorted configuration, there is no
intersection between the HL energy function. Therefore the Berry phase is
absent and the total spin $S$ is the same as for the case in which SO is
absent. However when the trimer is fixed in a symmetric configuration, the HL
gap vanishes for a few spin orientations in the plane of the clusters. Note
that, as in the ordinary static JT effect, the system lifts the degeneracy by
choosing a "distorted" configuration of the slow variables (away from the
"symmetric" one) that lowers the total energy. In $Co_3$ this configuration is
the spin orientation orthogonal to the trimer plane. An analysis of the HL
energy surfaces in the trimer plane reveals that there are 6 degeneracy points
separated by very shallow energy barriers. The fact that the Chern number of
the cluster changes from $S=5/2$ to $S=3/2$ when SO is included, seems to
indicate that paths that encircle an even {\it number} of conical intersections
have dominant tunneling rates.

\section{Conclusions and Outlook}

%
The crossover between macroscopic magnetic particles and atomic spins
passes through the rich and complex world of magnetic nano-clusters.
Interest in exploring this world is growing because of the push toward
denser magnetic information storage media and an ensuing interest in
achieving a deeper understanding of fundamental limits, and in identifying
potentially attractive systems.
From the classical point of view the most important property of a magnetic
particle is the dependence of its energy on moment orientation, {\em i.e.} its
magnetic anisotropy.  Magnetic anisotropy in magnetic nano-clusters is due
almost entirely to spin-orbit interactions which induce an orbital contribution
to the moment and therefore a dependence of energy on the orientation of the
moment relative to the spatial arrangement of atoms in the cluster.

In this paper
our interest has been focused on the quantum
effects which become important in small magnetic clusters.
%
%
In particular, we have explored a SDFT-based approach
to derive a {\em giant spin} low-energy effective Hamiltonian
from microscopic physics. This Hamiltonian is a quantum generalization of
the magnetic anisotropy energy, and describes a well-defined
group of low-lying energy levels which can be identified
with the magnetic moment orientation degree-of-freedom.

Our approach is orbital-based, rather than local-moment based, and is
intended to enable a practical description of systems in which the spins are
carried by orbitals which are widely spread over many of the cluster atoms.
Loosely speaking, our approach is intended to enable a quantum description of
metallic magnetic nano-clusters in which the spin-moment per atom in the absence of
spin-orbit interactions is not an integer.  The quantization rules of the
classical anisotropy energy are calculated by adding the adiabatic Berry phase
contributions from all itinerant orbitals which behave collectively
in the giant spin
approximation which is appropriate for slow (low-excitation energy) dynamics.
We believe that the approach we have taken will be accurate whenever
the fundamental
assumptions which underly the giant spin approximation are valid.

In our approach both the anisotropy energy functional and the Berry
curvature functional are calculated by solving the DFT Kohn-Sham equations
using standard spin-polarized density functional methods that include the spin-orbit interaction.
The average of the Berry curvature over magnetization directions, a
topological invariant known as Berry-phase Chern number, determines the total
angular momentum $J$ and reduces to the spin angular momentum $S$ in the absence of
spin-orbit interactions.  In our theory
the Chern number determines the dimension of the quantum giant-spin Hamiltonian.
We find that $J$ can deviate considerably from the total spin in the
absence of spin-orbit interactions $S$ when the HOMO-LUMO gap is small.
This is often the case in transition metal magnetic nano-clusters, which are often
characterized by a dense level structure near the Fermi level even when the clusters are quite small.
When the HL gap is small, the magnetization-dependent
Berry curvature has large fluctuations and its effect on
the effective spin Hamiltonian, which is usually neglected in treatments of molecular magnets,
is very significant. We call this effect topological since it is related
to the topological Berry-phase Chern number, and can be
intuitively described as a stretching
or a compression of the classical magnetic anisotropy landscape.
Although we have not performed calculations for these systems, we do not expect the
topological effect to be important in molecular magnets with well defined local moments.
For metals however, the Berry phase effect modifies qualitatively the
collective dynamics of the magnetization degree of freedom.  It is clear that
extremely large topological effects often signal a breakdown of the giant-spin
approximation.  In that case we view our approach as an attempt to construct the
best possible giant-spin model.

We have carried out our DFT-based quantum giant-spin model
construction procedure for dimers and for other clusters containing up to five
identical atoms.
We find that dimers can display a huge magnetic anisotropy because of their
strong axial symmetry.  For example we find anisotropies as large as $\approx
500$ K in Rh$_2$, which are accompanied by important topological effects that
widen the magnetic anisotropy barrier between the two equivalent uniaxial
minima. Both properties are closely related to the HOMO-LUMO degeneracy
mentioned above, and suggest that magnetic dimer systems could be of potential
technological importance for magnetic storage applications if they could be
embedded in an appropriate lattice. Such large magnetic anisotropies for such
small systems could represent the ultimate limit of magnetic storage
capability. Furthermore thicker energy barriers will tend to increase the
spin-relaxation time and suppress the microscopic quantum mechanical tunneling
of the magnetization. Larger clusters such as Co$_3$, Co$_4$, and Co$_5$,
typically manage to avoid quasi-degeneracy between the HOMO-LUMO levels by
means of Jahn-Teller distortions. In this case the impact of the topological
Berry curvature of the spin Hamiltonian is minor. If however a degeneracy if
forced on the system by imposing a particular symmetric configuration, the
singular Berry curvature once again plays a dominant role. Here we have shown
that the Berry curvature can even become negative and our procedure, strictly
speaking, breaks down, signaling that a description of the low-energy spectrum
in terms of an individual collective spin is not possible. For these cases one
is compelled to reintroduce the relevant electronic degrees of freedom involved
in the degeneracy at the Fermi level. We have sketched how this procedure
should be carried out in Sec.~\ref{alt_app}.

We have also looked at several hybrid clusters, in
particular at linear
Mn$_{x}$N$_{y}$-clusters in which the interatomic distances have
been locked to mimic the effect of the epitaxial registration on a
CuN surface. The study of this system is partly motivated by
recent STM experiments\cite{Hirjibehedin_sci06:_MnCuN}
that are able for the first time
to probe the collective spin dynamics of quantum engineered magnetic
clusters via inelastic electron tunneling. Experimentally these
clusters are found to order antiferromagnetically. We have investigated
this possibility and carried out our procedure for this situation.
We find that these systems show different topological effects
depending on the configuration studied, and exhibit high
sensitivity to the position of the N atoms coming from the CuN substrate.
The results offer
an alternative interpretation for the Mn dimer differential
conduction spectrum and in the case of the Mn trimer our findings
are compatible with the experimental results.
We should however point out that our calculations are for
free clusters only, namely we did not include any complicating effects
originating from the supporting CuN surface that are likely to
be important.

Our findings
indicate that powerful effects of non-trivial spin-space topology
are present in many small TM clusters, and cannot be ignored when
determining the collective spin dynamics. Furthermore, our approach
offers a way to extract the total spin of the system, without
assigning locally defined atomic spins. Only more detailed calculations
that include supporting surfaces and comparison with more accurate
experiments will tell if the effects predicted by our
theory can be observed in these systems.


\begin{thebibliography}{0}
\expandafter\ifx\csname natexlab\endcsname\relax\def\natexlab#1{#1}\fi
\expandafter\ifx\csname bibnamefont\endcsname\relax
  \def\bibnamefont#1{#1}\fi
\expandafter\ifx\csname bibfnamefont\endcsname\relax
  \def\bibfnamefont#1{#1}\fi
\expandafter\ifx\csname citenamefont\endcsname\relax
  \def\citenamefont#1{#1}\fi
\expandafter\ifx\csname url\endcsname\relax
  \def\url#1{\texttt{#1}}\fi
\expandafter\ifx\csname urlprefix\endcsname\relax\def\urlprefix{URL }\fi
\providecommand{\bibinfo}[2]{#2}
\providecommand{\eprint}[2][]{\url{#2}}

\end{thebibliography}


\begin{thebibliography}{53}
\expandafter\ifx\csname natexlab\endcsname\relax\def\natexlab#1{#1}\fi
\expandafter\ifx\csname bibnamefont\endcsname\relax
  \def\bibnamefont#1{#1}\fi
\expandafter\ifx\csname bibfnamefont\endcsname\relax
  \def\bibfnamefont#1{#1}\fi
\expandafter\ifx\csname citenamefont\endcsname\relax
  \def\citenamefont#1{#1}\fi
\expandafter\ifx\csname url\endcsname\relax
  \def\url#1{\texttt{#1}}\fi
\expandafter\ifx\csname urlprefix\endcsname\relax\def\urlprefix{URL }\fi
\providecommand{\bibinfo}[2]{#2} \providecommand{\eprint}[2][]{\url{#2}}

\bibitem[{\citenamefont{Sellmyer and Skomski}(2006)}]{skomski_book}
\bibinfo{editor}{\bibfnamefont{D.}~\bibnamefont{Sellmyer}} \bibnamefont{and}
  \bibinfo{editor}{\bibfnamefont{R.}~\bibnamefont{Skomski}}, eds.,
  \emph{\bibinfo{title}{Advanced Magnetic Nanostructures}}
  (\bibinfo{publisher}{Springer}, \bibinfo{address}{New York},
  \bibinfo{year}{2006}).

\bibitem[{\citenamefont{Gatteschi et~al.}(2006)\citenamefont{Gatteschi,
  Sessoli, and Villain}}]{molmagnet}
\bibinfo{author}{\bibfnamefont{D.}~\bibnamefont{Gatteschi}},
  \bibinfo{author}{\bibfnamefont{R.}~\bibnamefont{Sessoli}}, \bibnamefont{and}
  \bibinfo{author}{\bibfnamefont{J.}~\bibnamefont{Villain}},
  \emph{\bibinfo{title}{Molecular Nanomagnets}} (\bibinfo{publisher}{Oxford,
  New York}, \bibinfo{year}{2006}).

\bibitem[{\citenamefont{Sun et~al.}(2000)\citenamefont{Sun, Murray, Weller,
  Folks, and Moser}}]{murray_science2000}
\bibinfo{author}{\bibfnamefont{S.}~\bibnamefont{Sun}},
  \bibinfo{author}{\bibfnamefont{C.~B.} \bibnamefont{Murray}},
  \bibinfo{author}{\bibfnamefont{D.}~\bibnamefont{Weller}},
  \bibinfo{author}{\bibfnamefont{L.}~\bibnamefont{Folks}}, \bibnamefont{and}
  \bibinfo{author}{\bibfnamefont{A.}~\bibnamefont{Moser}},
  \bibinfo{journal}{Science} \textbf{\bibinfo{volume}{287}},
  \bibinfo{pages}{1989} (\bibinfo{year}{2000}).

\bibitem[{\citenamefont{F\'elix-Medina
  et~al.}(2003)\citenamefont{F\'elix-Medina, Dorantes-D\'avila, and
  Pastor}}]{pastor2003}
\bibinfo{author}{\bibfnamefont{R.}~\bibnamefont{F\'elix-Medina}},
  \bibinfo{author}{\bibfnamefont{J.}~\bibnamefont{Dorantes-D\'avila}},
  \bibnamefont{and} \bibinfo{author}{\bibfnamefont{G.~M.}
  \bibnamefont{Pastor}}, \bibinfo{journal}{Phys. Rev. B}
  \textbf{\bibinfo{volume}{67}}, \bibinfo{pages}{094430}
  (\bibinfo{year}{2003}).

\bibitem[{\citenamefont{Cehovin et~al.}(2002)\citenamefont{Cehovin, Canali, and
  MacDonald}}]{ac_cmc_ahm2002}
\bibinfo{author}{\bibfnamefont{A.}~\bibnamefont{Cehovin}},
  \bibinfo{author}{\bibfnamefont{C.~M.} \bibnamefont{Canali}},
  \bibnamefont{and} \bibinfo{author}{\bibfnamefont{A.~H.}
  \bibnamefont{MacDonald}}, \bibinfo{journal}{Phys. Rev. B}
  \textbf{\bibinfo{volume}{66}}, \bibinfo{pages}{094430}
  (\bibinfo{year}{2002}).

\bibitem[{\citenamefont{Skomski}(2003)}]{skomski_nanomag03}
\bibinfo{author}{\bibfnamefont{R.}~\bibnamefont{Skomski}}, \bibinfo{journal}{J.
  Phys.: Condens. Matter} \textbf{\bibinfo{volume}{15}}, \bibinfo{pages}{841}
  (\bibinfo{year}{2003}).

\bibitem[{\citenamefont{Eber et~al.}(2005)\citenamefont{Eber, Bornemann, Minar,
  Kosuth, Sipr, Dederichs, Zeller, and Cabri}}]{ebert05}
\bibinfo{author}{\bibfnamefont{H.}~\bibnamefont{Eber}},
  \bibinfo{author}{\bibfnamefont{S.}~\bibnamefont{Bornemann}},
  \bibinfo{author}{\bibfnamefont{J.}~\bibnamefont{Minar}},
  \bibinfo{author}{\bibfnamefont{M.}~\bibnamefont{Kosuth}},
  \bibinfo{author}{\bibfnamefont{O.}~\bibnamefont{Sipr}},
  \bibinfo{author}{\bibfnamefont{P.~H.} \bibnamefont{Dederichs}},
  \bibinfo{author}{\bibnamefont{Zeller}}, \bibnamefont{and}
  \bibinfo{author}{\bibfnamefont{I.}~\bibnamefont{Cabri}},
  \bibinfo{journal}{Phase Transitions} \textbf{\bibinfo{volume}{78}},
  \bibinfo{pages}{71} (\bibinfo{year}{2005}).

\bibitem[{\citenamefont{Kashyap et~al.}(2006)\citenamefont{Kashyap, Sabirianov,
  and Jaswal}}]{kashyap}
\bibinfo{author}{\bibfnamefont{A.}~\bibnamefont{Kashyap}},
  \bibinfo{author}{\bibfnamefont{R.}~\bibnamefont{Sabirianov}},
  \bibnamefont{and} \bibinfo{author}{\bibfnamefont{S.~S.}
  \bibnamefont{Jaswal}}, in \emph{\bibinfo{booktitle}{Advanced Magnetic
  Nanostructure}}, edited by
  \bibinfo{editor}{\bibfnamefont{D.}~\bibnamefont{Sellmyer}} \bibnamefont{and}
  \bibinfo{editor}{\bibfnamefont{R.}~\bibnamefont{Skomski}}
  (\bibinfo{publisher}{Springer}, \bibinfo{year}{2006}).

\bibitem[{\citenamefont{Billas et~al.}(1994)\citenamefont{Billas, Ch\^atelain,
  and de~Heer}}]{billas1994}
\bibinfo{author}{\bibfnamefont{I.~M.~L.} \bibnamefont{Billas}},
  \bibinfo{author}{\bibfnamefont{A.}~\bibnamefont{Ch\^atelain}},
  \bibnamefont{and} \bibinfo{author}{\bibfnamefont{W.~A.}
  \bibnamefont{de~Heer}}, \bibinfo{journal}{Science}
  \textbf{\bibinfo{volume}{265}}, \bibinfo{pages}{1682} (\bibinfo{year}{1994}).

\bibitem[{\citenamefont{Xu et~al.}(2005)\citenamefont{Xu, Yin, Moro, and
  de~Heer}}]{deHeer_prl05}
\bibinfo{author}{\bibfnamefont{X.}~\bibnamefont{Xu}},
  \bibinfo{author}{\bibfnamefont{S.}~\bibnamefont{Yin}},
  \bibinfo{author}{\bibfnamefont{R.}~\bibnamefont{Moro}}, \bibnamefont{and}
  \bibinfo{author}{\bibfnamefont{W.~A.} \bibnamefont{de~Heer}},
  \bibinfo{journal}{Phys. Rev. Lett} \textbf{\bibinfo{volume}{95}},
  \bibinfo{pages}{237209} (\bibinfo{year}{2005}).

\bibitem[{\citenamefont{Tiago et~al.}(2006)\citenamefont{Tiago, Zhou, Alemany,
  Saad, and Chelikowsky}}]{chelikowsky_prl06}
\bibinfo{author}{\bibfnamefont{M.~L.} \bibnamefont{Tiago}},
  \bibinfo{author}{\bibfnamefont{Y.}~\bibnamefont{Zhou}},
  \bibinfo{author}{\bibfnamefont{M.~M.~G.} \bibnamefont{Alemany}},
  \bibinfo{author}{\bibfnamefont{Y.}~\bibnamefont{Saad}}, \bibnamefont{and}
  \bibinfo{author}{\bibfnamefont{J.~R.} \bibnamefont{Chelikowsky}},
  \bibinfo{journal}{Phys.~Rev.~Lett,} \textbf{\bibinfo{volume}{97}},
  \bibinfo{pages}{147201} (\bibinfo{year}{2006}).

\bibitem[{\citenamefont{Jamet et~al.}(2001)\citenamefont{Jamet, Wernsdorfer,
  Thirion, Mailly, Dupuis, M\'elinon, and P\'eres}}]{jamet2001}
\bibinfo{author}{\bibfnamefont{M.}~\bibnamefont{Jamet}},
  \bibinfo{author}{\bibfnamefont{W.}~\bibnamefont{Wernsdorfer}},
  \bibinfo{author}{\bibfnamefont{C.}~\bibnamefont{Thirion}},
  \bibinfo{author}{\bibfnamefont{D.}~\bibnamefont{Mailly}},
  \bibinfo{author}{\bibfnamefont{V.}~\bibnamefont{Dupuis}},
  \bibinfo{author}{\bibfnamefont{P.}~\bibnamefont{M\'elinon}},
  \bibnamefont{and} \bibinfo{author}{\bibfnamefont{A.}~\bibnamefont{P\'eres}},
  \bibinfo{journal}{Phys. Rev. Lett.} \textbf{\bibinfo{volume}{86}},
  \bibinfo{pages}{4676} (\bibinfo{year}{2001}).

\bibitem[{\citenamefont{Gambarella et~al.}(2003)\citenamefont{Gambarella,
  Rusponi, Veronese, Dhesi, Grazioli, Dallmeyer, Cabria, Zeller, Dederichs,
  Kern et~al.}}]{gambarella03Sci}
\bibinfo{author}{\bibfnamefont{P.}~\bibnamefont{Gambarella}},
  \bibinfo{author}{\bibfnamefont{S.}~\bibnamefont{Rusponi}},
  \bibinfo{author}{\bibfnamefont{M.}~\bibnamefont{Veronese}},
  \bibinfo{author}{\bibfnamefont{S.~S.} \bibnamefont{Dhesi}},
  \bibinfo{author}{\bibfnamefont{C.}~\bibnamefont{Grazioli}},
  \bibinfo{author}{\bibfnamefont{A.}~\bibnamefont{Dallmeyer}},
  \bibinfo{author}{\bibfnamefont{I.}~\bibnamefont{Cabria}},
  \bibinfo{author}{\bibfnamefont{R.}~\bibnamefont{Zeller}},
  \bibinfo{author}{\bibfnamefont{P.~H.} \bibnamefont{Dederichs}},
  \bibinfo{author}{\bibfnamefont{K.}~\bibnamefont{Kern}}, \bibnamefont{et~al.},
  \bibinfo{journal}{Science} \textbf{\bibinfo{volume}{300}},
  \bibinfo{pages}{1130} (\bibinfo{year}{2003}).

\bibitem[{\citenamefont{Gambarella et~al.}(2002)\citenamefont{Gambarella,
  Dallmeyer, Maiti, Malagoli, amd K.~Kern, and Carbone}}]{gambarella02Nat1d}
\bibinfo{author}{\bibfnamefont{P.}~\bibnamefont{Gambarella}},
  \bibinfo{author}{\bibfnamefont{A.}~\bibnamefont{Dallmeyer}},
  \bibinfo{author}{\bibfnamefont{K.}~\bibnamefont{Maiti}},
  \bibinfo{author}{\bibfnamefont{M.~C.} \bibnamefont{Malagoli}},
  \bibinfo{author}{\bibfnamefont{W.~E.} \bibnamefont{amd K.~Kern}},
  \bibnamefont{and} \bibinfo{author}{\bibfnamefont{C.}~\bibnamefont{Carbone}},
  \bibinfo{journal}{Nature} \textbf{\bibinfo{volume}{416}},
  \bibinfo{pages}{301} (\bibinfo{year}{2002}).

\bibitem[{\citenamefont{Mokrousov et~al.}(2006)\citenamefont{Mokrousov,
  Bihlmayer, Heinze, and Bl{\"u}gel}}]{mokrousov_prl06}
\bibinfo{author}{\bibfnamefont{Y.}~\bibnamefont{Mokrousov}},
  \bibinfo{author}{\bibfnamefont{G.}~\bibnamefont{Bihlmayer}},
  \bibinfo{author}{\bibfnamefont{S.}~\bibnamefont{Heinze}}, \bibnamefont{and}
  \bibinfo{author}{\bibfnamefont{S.}~\bibnamefont{Bl{\"u}gel}},
  \bibinfo{journal}{Phys.~Rev.~Lett,} \textbf{\bibinfo{volume}{96}},
  \bibinfo{pages}{147201} (\bibinfo{year}{2006}).

\bibitem[{\citenamefont{Strandberg et~al.}(2007)\citenamefont{Strandberg,
  Canali, and MacDonald}}]{tos_cmc_ahm_natmat2007}
\bibinfo{author}{\bibfnamefont{T.~O.} \bibnamefont{Strandberg}},
  \bibinfo{author}{\bibfnamefont{C.~M.} \bibnamefont{Canali}},
  \bibnamefont{and} \bibinfo{author}{\bibfnamefont{A.~H.}
  \bibnamefont{MacDonald}}, \bibinfo{journal}{Nature Materials}
  \textbf{\bibinfo{volume}{6}}, \bibinfo{pages}{648} (\bibinfo{year}{2007}).

\bibitem[{\citenamefont{Bode et~al.}(2004)\citenamefont{Bode, Pietzsch,
  Kubetzka, and Wiesendanger}}]{bode_prl04}
\bibinfo{author}{\bibfnamefont{M.}~\bibnamefont{Bode}},
  \bibinfo{author}{\bibfnamefont{O.}~\bibnamefont{Pietzsch}},
  \bibinfo{author}{\bibfnamefont{A.}~\bibnamefont{Kubetzka}}, \bibnamefont{and}
  \bibinfo{author}{\bibfnamefont{R.}~\bibnamefont{Wiesendanger}},
  \bibinfo{journal}{Phys.~Rev.~Lett,} \textbf{\bibinfo{volume}{92}},
  \bibinfo{pages}{067201} (\bibinfo{year}{2004}).

\bibitem[{\citenamefont{Gunther and Barbara}(1995)}]{qtm94}
\bibinfo{editor}{\bibfnamefont{L.}~\bibnamefont{Gunther}} \bibnamefont{and}
  \bibinfo{editor}{\bibfnamefont{B.}~\bibnamefont{Barbara}}, eds.,
  \emph{\bibinfo{title}{Quantum Tunneling of Magnetization}}, QTM´94
  (\bibinfo{publisher}{Kluwer, Dordrecht}, \bibinfo{year}{1995}).

\bibitem[{\citenamefont{Chudnovsky and Tejada}(1998)}]{chudnovsky_tejada98}
\bibinfo{author}{\bibfnamefont{E.}~\bibnamefont{Chudnovsky}} \bibnamefont{and}
  \bibinfo{author}{\bibfnamefont{J.}~\bibnamefont{Tejada}},
  \emph{\bibinfo{title}{Macroscopic Quantum Tunneling of the Magnetic Moment}}
  (\bibinfo{publisher}{Cambridge University Press},
  \bibinfo{address}{Cambridge}, \bibinfo{year}{1998}).

\bibitem[{\citenamefont{Hirjibehedin et~al.}(2006)\citenamefont{Hirjibehedin,
  Lutz, and Heinrich}}]{Hirjibehedin_sci06:_MnCuN}
\bibinfo{author}{\bibfnamefont{C.~F.} \bibnamefont{Hirjibehedin}},
  \bibinfo{author}{\bibfnamefont{C.~P.} \bibnamefont{Lutz}}, \bibnamefont{and}
  \bibinfo{author}{\bibfnamefont{A.~J.} \bibnamefont{Heinrich}},
  \bibinfo{journal}{Science} \textbf{\bibinfo{volume}{312}},
  \bibinfo{pages}{1021} (\bibinfo{year}{2006}).

\bibitem[{\citenamefont{Canali et~al.}(2003)\citenamefont{Canali, Cehovin, and
  MacDonald}}]{ccmPRL03}
\bibinfo{author}{\bibfnamefont{C.~M.} \bibnamefont{Canali}},
  \bibinfo{author}{\bibfnamefont{A.}~\bibnamefont{Cehovin}}, \bibnamefont{and}
  \bibinfo{author}{\bibfnamefont{A.~H.} \bibnamefont{MacDonald}},
  \bibinfo{journal}{Phys. Rev. Lett.} \textbf{\bibinfo{volume}{91}},
  \bibinfo{pages}{046805} (\bibinfo{year}{2003}).

\bibitem[{\citenamefont{Berry}(1984)}]{originalberry_prs84}
\bibinfo{author}{\bibfnamefont{M.~V.} \bibnamefont{Berry}},
  \bibinfo{journal}{Proc. R. Soc. A} \textbf{\bibinfo{volume}{392}}
  (\bibinfo{year}{1984}).

\bibitem[{\citenamefont{Resta}(2000)}]{resta2000}
\bibinfo{author}{\bibfnamefont{R.}~\bibnamefont{Resta}}, \bibinfo{journal}{J.
  Phys.: Condens. Matter} \textbf{\bibinfo{volume}{12}}, \bibinfo{pages}{107}
  (\bibinfo{year}{2000}).

\bibitem[{\citenamefont{Auerbach}(1994)}]{auerbach94:_inter_elect_quant_magnet}
\bibinfo{author}{\bibfnamefont{A.}~\bibnamefont{Auerbach}},
  \emph{\bibinfo{title}{Interacting Electron and Quantum Magnetism}}
  (\bibinfo{publisher}{Springer-Verlag}, \bibinfo{address}{New York},
  \bibinfo{year}{1994}).

\bibitem[{\citenamefont{Bohm et~al.}(2003)\citenamefont{Bohm, Mostafazadeh,
  Koizumi, Niu, and Zwanziger}}]{qian_niu_book}
\bibinfo{author}{\bibfnamefont{A.}~\bibnamefont{Bohm}},
  \bibinfo{author}{\bibfnamefont{A.}~\bibnamefont{Mostafazadeh}},
  \bibinfo{author}{\bibfnamefont{H.}~\bibnamefont{Koizumi}},
  \bibinfo{author}{\bibfnamefont{Q.}~\bibnamefont{Niu}}, \bibnamefont{and}
  \bibinfo{author}{\bibfnamefont{J.}~\bibnamefont{Zwanziger}},
  \emph{\bibinfo{title}{The Geometric Phase in Quantum Systems}}
  (\bibinfo{publisher}{Springer-Verlag}, \bibinfo{address}{New York},
  \bibinfo{year}{2003}).

\bibitem[{\citenamefont{Cehovin et~al.}(2003)\citenamefont{Cehovin, Canali, and
  MacDonald}}]{cmc_ac_ahm2002pap4}
\bibinfo{author}{\bibfnamefont{A.}~\bibnamefont{Cehovin}},
  \bibinfo{author}{\bibfnamefont{C.~M.} \bibnamefont{Canali}},
  \bibnamefont{and} \bibinfo{author}{\bibfnamefont{A.~H.}
  \bibnamefont{MacDonald}}, \bibinfo{journal}{Phys. Rev. B}
  \textbf{\bibinfo{volume}{68}}, \bibinfo{pages}{014423}
  (\bibinfo{year}{2003}).

\bibitem[{\citenamefont{Uhl et~al.}(1994)\citenamefont{Uhl, Sandratski, and
  Kuber}}]{uhl1994}
\bibinfo{author}{\bibfnamefont{M.}~\bibnamefont{Uhl}},
  \bibinfo{author}{\bibfnamefont{L.~M.} \bibnamefont{Sandratski}},
  \bibnamefont{and} \bibinfo{author}{\bibfnamefont{J.}~\bibnamefont{Kuber}},
  \bibinfo{journal}{Phys. Rev. B} \textbf{\bibinfo{volume}{50}},
  \bibinfo{pages}{291} (\bibinfo{year}{1994}).

\bibitem[{\citenamefont{Niu and Kleinman}(1998)}]{niu1998}
\bibinfo{author}{\bibfnamefont{Q.}~\bibnamefont{Niu}} \bibnamefont{and}
  \bibinfo{author}{\bibfnamefont{L.}~\bibnamefont{Kleinman}},
  \bibinfo{journal}{Phys. Rev. Lett.} \textbf{\bibinfo{volume}{80}},
  \bibinfo{pages}{2205} (\bibinfo{year}{1998}).

\bibitem[{\citenamefont{Niu et~al.}(1999)\citenamefont{Niu, Wang, Kleinman,
  Liu, Nicholson, and Stocks}}]{niu1999}
\bibinfo{author}{\bibfnamefont{Q.}~\bibnamefont{Niu}},
  \bibinfo{author}{\bibfnamefont{X.}~\bibnamefont{Wang}},
  \bibinfo{author}{\bibfnamefont{L.}~\bibnamefont{Kleinman}},
  \bibinfo{author}{\bibfnamefont{W.~M.} \bibnamefont{Liu}},
  \bibinfo{author}{\bibfnamefont{D.~M.~C.} \bibnamefont{Nicholson}},
  \bibnamefont{and} \bibinfo{author}{\bibfnamefont{G.~M.}
  \bibnamefont{Stocks}}, \bibinfo{journal}{Phys. Rev. Lett.}
  \textbf{\bibinfo{volume}{83}}, \bibinfo{pages}{207} (\bibinfo{year}{1999}).

\bibitem[{\citenamefont{Gebauer and Baroni}(2000)}]{gebauer2000}
\bibinfo{author}{\bibfnamefont{R.}~\bibnamefont{Gebauer}} \bibnamefont{and}
  \bibinfo{author}{\bibfnamefont{S.}~\bibnamefont{Baroni}},
  \bibinfo{journal}{Phys. Rev. B} \textbf{\bibinfo{volume}{61}},
  \bibinfo{pages}{6459} (\bibinfo{year}{2000}).

\bibitem[{\citenamefont{Bylander et~al.}(2000)\citenamefont{Bylander, Niu, and
  Kleinman}}]{bylander2000}
\bibinfo{author}{\bibfnamefont{D.~M.} \bibnamefont{Bylander}},
  \bibinfo{author}{\bibfnamefont{Q.}~\bibnamefont{Niu}}, \bibnamefont{and}
  \bibinfo{author}{\bibfnamefont{L.}~\bibnamefont{Kleinman}},
  \bibinfo{journal}{Phys. Rev. B} \textbf{\bibinfo{volume}{61}},
  \bibinfo{pages}{11875} (\bibinfo{year}{2000}).

\bibitem[{\citenamefont{Blochl}(1994)}]{originalpaw_prb94}
\bibinfo{author}{\bibfnamefont{P.~E.} \bibnamefont{Blochl}},
  \bibinfo{journal}{Phys. Rev. B} \textbf{\bibinfo{volume}{50}},
  \bibinfo{pages}{17953} (\bibinfo{year}{1994}).

\bibitem[{\citenamefont{Kresse and Furthmuller}(1996)}]{vaspcode}
\bibinfo{author}{\bibfnamefont{G.}~\bibnamefont{Kresse}} \bibnamefont{and}
  \bibinfo{author}{\bibfnamefont{J.}~\bibnamefont{Furthmuller}},
  \bibinfo{journal}{Comput. Phys. Commun.} \textbf{\bibinfo{volume}{6}}
  (\bibinfo{year}{1996}).

\bibitem[{\citenamefont{Kresse and Joubert}(1999)}]{vasppaw_psps}
\bibinfo{author}{\bibfnamefont{G.}~\bibnamefont{Kresse}} \bibnamefont{and}
  \bibinfo{author}{\bibfnamefont{J.}~\bibnamefont{Joubert}},
  \bibinfo{journal}{Phys. Rev. B} \textbf{\bibinfo{volume}{59}},
  \bibinfo{pages}{1758} (\bibinfo{year}{1999}).

\bibitem[{\citenamefont{Vosko et~al.}(1980)\citenamefont{Vosko, Wilk, and
  Nusair}}]{vosko-wilk-nusair_param}
\bibinfo{author}{\bibfnamefont{S.~H.} \bibnamefont{Vosko}},
  \bibinfo{author}{\bibfnamefont{L.}~\bibnamefont{Wilk}}, \bibnamefont{and}
  \bibinfo{author}{\bibfnamefont{M.}~\bibnamefont{Nusair}},
  \bibinfo{journal}{Can J. Phys.} \textbf{\bibinfo{volume}{58}},
  \bibinfo{pages}{1200} (\bibinfo{year}{1980}).

\bibitem[{\citenamefont{Hobbs et~al.}(2000)\citenamefont{Hobbs, Kresse, and
  Hafner}}]{noncol-vasp}
\bibinfo{author}{\bibfnamefont{D.}~\bibnamefont{Hobbs}},
  \bibinfo{author}{\bibfnamefont{G.}~\bibnamefont{Kresse}}, \bibnamefont{and}
  \bibinfo{author}{\bibfnamefont{J.}~\bibnamefont{Hafner}},
  \bibinfo{journal}{Phys. Rev. B} \textbf{\bibinfo{volume}{62}},
  \bibinfo{pages}{11556} (\bibinfo{year}{2000}).

\bibitem[{\citenamefont{Pederson and Jackson}(1991)}]{pederson_prb91:Pfrac_occ}
\bibinfo{author}{\bibfnamefont{M.~R.} \bibnamefont{Pederson}} \bibnamefont{and}
  \bibinfo{author}{\bibfnamefont{K.~A.} \bibnamefont{Jackson}},
  \bibinfo{journal}{Phys. Rev. B} \textbf{\bibinfo{volume}{43}},
  \bibinfo{pages}{7312} (\bibinfo{year}{1991}).

\bibitem[{\citenamefont{Jamorski et~al.}(1997)\citenamefont{Jamorski, Martinez,
  Castro, and Salahub}}]{castro_prb97:_smearquote_and_coclusters}
\bibinfo{author}{\bibfnamefont{C.}~\bibnamefont{Jamorski}},
  \bibinfo{author}{\bibfnamefont{A.}~\bibnamefont{Martinez}},
  \bibinfo{author}{\bibfnamefont{M.}~\bibnamefont{Castro}}, \bibnamefont{and}
  \bibinfo{author}{\bibfnamefont{D.~R.} \bibnamefont{Salahub}},
  \bibinfo{journal}{Phys. Rev. B} \textbf{\bibinfo{volume}{55}},
  \bibinfo{pages}{10905} (\bibinfo{year}{1997}).

\bibitem[{\citenamefont{Micherlini et~al.}(1998)\citenamefont{Micherlini, Diez,
  and Jubert}}]{smearquote2_ni}
\bibinfo{author}{\bibfnamefont{M.~C.} \bibnamefont{Micherlini}},
  \bibinfo{author}{\bibfnamefont{R.~P.} \bibnamefont{Diez}}, \bibnamefont{and}
  \bibinfo{author}{\bibfnamefont{A.~H.} \bibnamefont{Jubert}},
  \bibinfo{journal}{Int. J. of Quantum Chem.} \textbf{\bibinfo{volume}{70}},
  \bibinfo{pages}{699} (\bibinfo{year}{1998}).

\bibitem[{\citenamefont{Koelling and
  Harmon}(1977)}]{Koelling_jpc77:_scal_rel_soi}
\bibinfo{author}{\bibfnamefont{D.~D.} \bibnamefont{Koelling}} \bibnamefont{and}
  \bibinfo{author}{\bibfnamefont{B.~N.} \bibnamefont{Harmon}},
  \bibinfo{journal}{J. Phys. C} \textbf{\bibinfo{volume}{10}},
  \bibinfo{pages}{3107} (\bibinfo{year}{1977}).

\bibitem[{\citenamefont{MacDonald et~al.}(1980)\citenamefont{MacDonald,
  Pickett, and Koelling}}]{macdonald_jpc80:_scal_rel_soi}
\bibinfo{author}{\bibfnamefont{A.~H.} \bibnamefont{MacDonald}},
  \bibinfo{author}{\bibfnamefont{W.~E.} \bibnamefont{Pickett}},
  \bibnamefont{and} \bibinfo{author}{\bibfnamefont{D.~D.}
  \bibnamefont{Koelling}}, \bibinfo{journal}{J. Phys. C}
  \textbf{\bibinfo{volume}{13}}, \bibinfo{pages}{2675} (\bibinfo{year}{1980}).

\bibitem[{\citenamefont{Carrier and Wei}(2004)}]{soi-description}
\bibinfo{author}{\bibfnamefont{P.}~\bibnamefont{Carrier}} \bibnamefont{and}
  \bibinfo{author}{\bibfnamefont{S.-H.} \bibnamefont{Wei}},
  \bibinfo{journal}{Phys. Rev. B} \textbf{\bibinfo{volume}{70}},
  \bibinfo{pages}{035212} (\bibinfo{year}{2004}).

\bibitem[{\citenamefont{Fuller}(1981)}]{buckminsterfuller}
\bibinfo{author}{\bibnamefont{Fuller}}, \emph{\bibinfo{title}{Critical Path}}
  (\bibinfo{publisher}{St. Martin's Press}, \bibinfo{address}{New York},
  \bibinfo{year}{1981}).

\bibitem[{\citenamefont{Castro et~al.}(1997)\citenamefont{Castro, Jamorski, and
  Salahub}}]{castro_cpl97:_fenico}
\bibinfo{author}{\bibfnamefont{M.}~\bibnamefont{Castro}},
  \bibinfo{author}{\bibfnamefont{C.}~\bibnamefont{Jamorski}}, \bibnamefont{and}
  \bibinfo{author}{\bibfnamefont{D.~R.} \bibnamefont{Salahub}},
  \bibinfo{journal}{Chem. Phys. Lett.} \textbf{\bibinfo{volume}{271}},
  \bibinfo{pages}{133} (\bibinfo{year}{1997}).

\bibitem[{\citenamefont{Futschek et~al.}(2005)\citenamefont{Futschek, Marsman,
  and Hafner}}]{hafner_jpc05:_pd_rh}
\bibinfo{author}{\bibfnamefont{T.}~\bibnamefont{Futschek}},
  \bibinfo{author}{\bibfnamefont{M.}~\bibnamefont{Marsman}}, \bibnamefont{and}
  \bibinfo{author}{\bibfnamefont{J.}~\bibnamefont{Hafner}},
  \bibinfo{journal}{J. Phys.: Condens. Matter} \textbf{\bibinfo{volume}{17}},
  \bibinfo{pages}{5927} (\bibinfo{year}{2005}).

\bibitem[{\citenamefont{Herzberg}(1950)}]{herzberg}
\bibinfo{author}{\bibfnamefont{G.}~\bibnamefont{Herzberg}},
  \emph{\bibinfo{title}{Molecular Spectra and Molecular Structure}}
  (\bibinfo{publisher}{Van Nostrand Reinhold}, \bibinfo{address}{New York},
  \bibinfo{year}{1950}).

\bibitem[{\citenamefont{Kortus et~al.}(2002)\citenamefont{Kortus, Baruah, and
  Pederson}}]{mark_apl02:_cofe}
\bibinfo{author}{\bibfnamefont{J.}~\bibnamefont{Kortus}},
  \bibinfo{author}{\bibfnamefont{T.}~\bibnamefont{Baruah}}, \bibnamefont{and}
  \bibinfo{author}{\bibfnamefont{M.~R.} \bibnamefont{Pederson}},
  \bibinfo{journal}{Appl. Phys. Lett.} \textbf{\bibinfo{volume}{80}},
  \bibinfo{pages}{4193} (\bibinfo{year}{2002}).

\bibitem[{\citenamefont{Rao and Jena}(2002)}]{rao_prl02:_MnN}
\bibinfo{author}{\bibfnamefont{B.~K.} \bibnamefont{Rao}} \bibnamefont{and}
  \bibinfo{author}{\bibfnamefont{P.}~\bibnamefont{Jena}},
  \bibinfo{journal}{Phys. Rev. Lett.} \textbf{\bibinfo{volume}{89}},
  \bibinfo{pages}{185504} (\bibinfo{year}{2002}).

\bibitem[{\citenamefont{Bersuker}(2006)}]{bersuker_book06}
\bibinfo{author}{\bibfnamefont{I.}~\bibnamefont{Bersuker}},
  \emph{\bibinfo{title}{The Jahn-Teller Effect}} (\bibinfo{publisher}{Cambridge
  University Press, Cambridge, England}, \bibinfo{year}{2006}).

\bibitem[{\citenamefont{Shapere and Wilczek}(1989)}]{wilzek_book89}
\bibinfo{author}{\bibfnamefont{A.}~\bibnamefont{Shapere}} \bibnamefont{and}
  \bibinfo{author}{\bibfnamefont{F.}~\bibnamefont{Wilczek}},
  \emph{\bibinfo{title}{Geometric Phases in Physics}}
  (\bibinfo{publisher}{World Scientific}, \bibinfo{year}{1989}).

\bibitem[{\citenamefont{Garcia-Fernandez
  et~al.}(2006)\citenamefont{Garcia-Fernandez, Bersuker, and
  Boggs}}]{bersuker_prl06}
\bibinfo{author}{\bibfnamefont{P.}~\bibnamefont{Garcia-Fernandez}},
  \bibinfo{author}{\bibfnamefont{I.~B.} \bibnamefont{Bersuker}},
  \bibnamefont{and} \bibinfo{author}{\bibfnamefont{J.~E.} \bibnamefont{Boggs}},
  \bibinfo{journal}{Phys.~Rev.~Lett.} \textbf{\bibinfo{volume}{96}},
  \bibinfo{pages}{163005} (\bibinfo{year}{2006}).

\bibitem[{\citenamefont{Zwanzinger and Grant}(1987)}]{zwanzinger_87}
\bibinfo{author}{\bibfnamefont{J.}~\bibnamefont{Zwanzinger}} \bibnamefont{and}
  \bibinfo{author}{\bibfnamefont{E.}~\bibnamefont{Grant}}, \bibinfo{journal}{J.
  Chem Phys.} \textbf{\bibinfo{volume}{87}}, \bibinfo{pages}{2954}
  (\bibinfo{year}{1987}).

\bibitem[{\citenamefont{Koizumi and Bersuker}(1999)}]{koizumi_bersuker_prl99}
\bibinfo{author}{\bibfnamefont{H.}~\bibnamefont{Koizumi}} \bibnamefont{and}
  \bibinfo{author}{\bibfnamefont{I.~B.} \bibnamefont{Bersuker}},
  \bibinfo{journal}{Phys.~Rev.~Lett.} \textbf{\bibinfo{volume}{83}},
  \bibinfo{pages}{3009} (\bibinfo{year}{1999}).

\end{thebibliography}
\end{document}